\documentclass[useAMS,usenatbib]{mn2e}

\usepackage{graphicx}
\usepackage{gensymb}
\usepackage{natbib}
\usepackage{amssymb}
\usepackage{upgreek}
\usepackage{array}
\usepackage{multirow}
\usepackage[usenames]{xcolor}

\title[Radio observations of $\gamma$-ray-loud pulsars]{Constraints on viewing geometries from radio observations of $\gamma$-ray-loud pulsars using a novel method}
\author[S. C. Rookyard, P. Weltevrede and S. Johnston]{S. C. Rookyard$^{1}$\thanks{E-mail:
simon.rookyard@postgrad.manchester.ac.uk}, P. Weltevrede$^{1}$ and S. Johnston$^{2}$\\
$^{1}$Jodrell Bank Centre for Astrophysics, School of Physics and Astronomy, University of Manchester, Manchester M13 9PL, UK\\
$^{2}$CSIRO Astronomy and Space Science, Australia Telescope National Facility, Epping NSW 1710, Australia}
\begin{document}

\date{Accepted 1988 December 15. Received 1988 December 14; in original form 1988 October 11}

\pagerange{\pageref{firstpage}--\pageref{lastpage}} \pubyear{2002}

\maketitle

\label{firstpage}

\begin{abstract}
We present radio intensity and polarisation profiles of 
28 $\gamma$-ray-detected pulsars with the aim of putting 
constraints on their viewing geometries using data from the Parkes telescope.
Constraints are formed both from the goodness-of-fit of the position angles
to the Rotating Vector Model and from the beam 
opening angle considering aberration and retardation effects. Uncertainties on 
the relevant parameters are systematically taken into account in order to produce a more robust constraint, using a new approach. 
Surprisingly, we find that the distribution of the magnetic 
inclination angle ($\alpha$) in this subset of pulsars peaks at low 
values, contrary to the predictions of $\gamma$-ray models. We find a lack of correlation between these and a set of $\alpha$ values which were derived using $\gamma$-ray light curves, suggesting a problem in the interpretation of the data in one or both of these domains.
Finally, we also show that the $\alpha$ distribution of pulsars with 
multiple radio components is no different to that of single-component pulsars.

\end{abstract}


\section{Introduction}

The \emph{FERMI} satellite was launched on 2008 June 11 and has since 
greatly increased the number of known pulsars emitting in the $\gamma$-ray 
band. The principle instrument aboard the satellite is the Large 
Area Telescope (LAT; \citealt{aaa+09}), a pair-production telescope which 
is sensitive to energies in the range 20~MeV - 300~GeV. The impact of 
the mission so far can be seen from the 2nd \emph{FERMI} Pulsar Catalog 
\citep{aaa+13}, which details 117 detections (at $\geq$~100~MeV) 
of $\gamma$-ray-loud pulsars. This factor of 20 increase 
in the number of pulsars which can be studied in the high energy 
regime makes investigation of the pulsar $\gamma$-ray emission process 
more important than ever. 

Knowledge of the `viewing geometry' of these pulsars can significantly aid this 
study. Particularly important is the determination of the path traced 
across the magnetosphere by the observer's line of sight 
as the pulsar rotates, which can be characterised by two angles; 
the angle between the rotation and magnetic axes, $\alpha$, and the 
angle $\beta$ between the magnetic axis and the line of sight 
at the closest approach between the two. If these are known for 
a particular pulsar, a given model 
of the sky pattern of $\gamma$-ray emission (such as those described in \cite{wrw+09, rw10}) will be able to predict, for example, the 
shape of that pulsar's intensity profile. This can then be compared 
with observations, allowing the veracity of the model 
to be judged. The determination of the viewing geometry is therefore 
useful in the study of the pulsar emission mechanism. 

A recent study by \cite{phg+14} highlighted a general 
problem that $\gamma$-ray light-curves can only be meaningfully
interpreted when radio data are also considered. These authors used
$\gamma$-ray light-curve and radio profile fitting to constrain the
 viewing geometry in terms of $\alpha$ and
$\zeta=\alpha+\beta$, the angle between the line of sight and the 
rotation axis. They note that if they consider radio-loud pulsars, but
ignore their radio profiles in the fitting procedure, the resulting
$\beta$ values do not appear to be confined to small values. This is
unlikely to be realistic, given the fact that the radio emission
should be significantly more beamed compared to the
$\gamma$-rays. They conclude that $\gamma$-ray light-curve fitting
in general does not lead to unique solutions in $\alpha$ and $\zeta$
space and that radio observations are crucial to break this degeneracy.

The intention of this paper is to derive emission
geometries, which can facilitate the interpretation of $\gamma$-ray
light-curves and therefore potentially impose important constraints on
various proposed $\gamma$-ray models. Using data obtained with the 
Parkes radio telescope as part of the \emph{FERMI} timing programme 
\citep{wjm+10} we attempt to place constraints on the viewing geometry 
for 28 young, $\gamma$-ray-detected pulsars, all of which were included in 
\cite{aaa+13}. As the angles $\alpha$ and $\beta$ should be independent 
of observing frequency\footnote{It has been noted that scattering of the pulsed emission by the interstellar medium can, in some cases, lead to estimates of these angles which are sensitive to the observing frequency (see \cite{kj08} and $\S$ \ref{J0908}). However, the intrinsic values are frequency-independent and the uncertainty of a few degrees introduced by scattering is in general not the most significant source of uncertainty in the sample considered in this paper.}, the results derived 
here can be applied to observations of these pulsars at any part of 
the electromagnetic spectrum. A further point which is worthy of note is that, although there are large
uncertainties on most of the $\alpha$ values presented in this paper, $\beta$ will be shown to be well constrained for most of the sample. This is useful for attempts to constrain the $\gamma$-ray models as requiring a small value of $\beta$ greatly restricts the region of ($\alpha$, $\zeta$) parameter space which is of interest, helping to lift the aforementioned degeneracy.

One method of constraining $\alpha$ and $\beta$ is to fit the 
Rotating Vector Model (RVM; \citealt{rc69a}) to the observed polarisation
position angle (PA) swing. The PA of the linearly polarised 
emission, $\psi$, is a function of $\alpha$, $\beta$ and the pulse phase, 
$\phi$, according to 

\begin{equation}
\label{EqPA} \tan(\psi - \psi_0)=\frac{\sin(\phi - \phi_0) \sin\alpha}{\sin(\alpha + \beta) \cos\alpha - \cos(\alpha + \beta) \sin\alpha \cos(\phi - \phi_0)} , 
\end{equation}
where $\psi_0$ and $\phi_0$ are the position angle and pulse phase, 
respectively, of the steepest part of this curve \citep{kom70}. 
The steepest gradient of the RVM curve is given by 

\begin{equation}
\label{EqdPAdPhi} \left(\frac{d\psi}{d\phi}\right)_{\mathrm{MAX}}=\frac{\sin\alpha}{\sin\beta}
\end{equation}
\citep{kom70}. If the observed PA swing shows little curvature,
only its gradient is constrained, leading to a constraint on 
$\sin\alpha$/$\sin\beta$ (see $\S$~\ref{SectChi2Surfaces} for details 
of the fitting procedure). In ($\alpha$, $\beta$) space this results 
in a characteristic `banana' shape of the goodness-of-fit, which can be seen in, for example, Fig.~\ref{FigJ0631+1036}.

The steepest gradient occurs at $\phi_0$, the `inflection point' of the PA curve. The RVM predicts that this coincides with the passage of the line 
of sight through the `fiducial plane', the plane 
containing both the rotation and magnetic axes.
However, in practice observations show the inflection point to be delayed relative to the position of the fiducial plane as inferred from the intensity profile ($\phi_{\mathrm{fid}}$) by an amount of rotational phase $\Delta\phi = \phi_0 - \phi_{\mathrm{fid}}$. This delay is predicted by relativistic effects known as aberration and retardation (A/R). The net delay predicted by A/R effects is

\begin{equation}
\label{EqOffset} \Delta\phi = \frac{8\pi h_{\mathrm{em}}}{Pc}   ,
\end{equation}
where $P$ is the rotation period of the star, $c$ is the speed of light and 
$h_{\mathrm{em}}$ is the emission height, the distance of the emission 
region from the centre 
of the star \citep{bcw91, dzg05}. The observed delay can be shown to be independent of $\alpha$ and $\beta$ \citep{drh04}. Estimates can be obtained for $\phi_0$ from RVM fitting (see $\S$~\ref{SectChi2Surfaces}) and for $\phi_{\mathrm{fid}}$ based on the profile morphology (see $\S$~\ref{SectEmissionHeightConstraint}). Given these estimates $\Delta\phi$ can be determined from the data and so Eq.~\ref{EqOffset} can be used to calculate $h_{\mathrm{em}}$. As discussed by for instance \cite{kj07}, variations in the emission height between different components will cause distortions in the PA curve and hence could affect the determination of $\Delta\phi$. This effect is probably strongest for pulsars older than those discussed in this paper, hence we assume $h_{\mathrm{em}}$ to be constant across the emission region for a given pulsar. The potential effects of variations in the emission height will be considered in \cite{rwj14b}.

The angles $\alpha$ and $\beta$ are also related to another observable, the pulse width of the radio profile. If the radio beam is defined 
as the cone bounded by tangents to the last open field lines 
at the emission height, the half-opening angle of the beam, $\rho$, can be calculated. Assuming that the magnetic field is dipolar,

\begin{equation}
\label{EqBeamHalfOpeningAngle} \rho = \theta_{\mathrm{PC}} + \arctan\left(\frac{1}{2} \tan\theta_{\mathrm{PC}}\right)   ,
\end{equation}
with $\theta_{\mathrm{PC}}$, the angular radius of the open-field-line region, 
given by
\begin{equation}
\label{EqThetaPC} \theta_{\mathrm{PC}} = \arcsin\left(\sqrt{\frac{2\pi h_{\mathrm{em}}}{Pc}}\right)   
\end{equation}
(e.g., \citealt{PulsarAstronomy}).

\cite{ggr84} showed that the half-opening angle of a conical emission 
beam centred on the magnetic axis can be related to $W_{\mathrm{open}}$, the range 
in pulse phase for which the line of sight samples the open-field-line region. 
This relation is dependent on the viewing geometry via the equation

\begin{equation}
\label{EqRhoContours} \cos\rho = \cos\alpha\cos(\alpha + \beta) + \sin\alpha\sin(\alpha + \beta)\cos\left(\frac{W_{\mathrm{open}}}{2}\right).
\end{equation}

A similar method of combining this information has been used by previous 
authors (e.g., \citealt{jw06, waa+10}). However, this work differs from 
those in two important ways. Firstly, we consider uncertainties on the 
relevant parameters, which allows us to determine errorbars on the fit 
parameters in a more objective way. Secondly, although we use the core-cone 
model \citep{ran93} as a basis for our estimation of $\phi_{\mathrm{fid}}$, 
we argue that the conservative nature of these estimates means that 
our results are unlikely to be inconsistent with the patchy beams suggested 
by \cite{lm88} (see $\S$~\ref{SectEmissionHeightConstraint}). 

The paper is organised as follows: in $\S$~\ref{SectMethod} 
we describe the observations and methodology used. In $\S$~\ref{SectResults} 
we present intensity and polarisation profiles and viewing geometry 
constraints for the individual pulsars and a table 
with constraints is compiled. In $\S$~\ref{SectAlphaDistribution} 
we discuss the derived distributions of magnetic 
inclination angles for pulsars exhibiting single-component and multiple-component profiles. We also compare our results with those obtained from $\gamma$-ray modeling by \cite{phg+14}. \cite{rwj14b} will discuss the overall distribution of $\alpha$ values, and its dependence on assumptions about the radio beam.

\section{Observations and Analysis}
\label{SectMethod}

\subsection{Observations}

The data used were collected as part of the \emph{FERMI} timing 
programme at the Parkes radio telescope, described in detail by \cite{wjm+10}.
A total of 168 pulsars are observed drawn from a list of pulsars with large-spin-down rates, and hence potentially high energy $\gamma$-ray emission, given by 
\cite{sgc+08} supplemented by a few additional objects.
The programme consists of monthly observations
at 1369 MHz (20 cm) with a 256 MHz bandwidth and twice-annual 
observations of the same pulsars at 3100 MHz (10 cm) and 685 MHz 
(50 cm), with bandwidths of 1024 MHz and 40 MHz respectively \citep{pkj+13}. 
At each frequency the band is divided into 1024 channels and the 
pulse phase resolution is 1024 bins per pulse.

In this paper we concentrate on the sample of 28 pulsars which are 
$\gamma$-ray-loud and published in the second Fermi pulsar 
catalogue \citep{aaa+13}.
We use observations taken between 2007 April and 2013 October. For the majority
of pulsars, the observations at 1369~MHz were used. Some pulsars however are
significantly scatter-broadened due to propagation through the interstellar medium at this frequency and so the observations at 3100~MHz were used instead (see Table~\ref{TabDerivedParameters}). In the cases of PSRs J1019--5749 and J1410--6132 the profile is still severely scatter-broadened at 3100~MHz; in the latter case we used unpublished Parkes data at 6100~MHz.
The data were polarisation calibrated and individual observations of the same pulsar were summed largely following the method described in \cite{wj08b}, resulting in a single profile for each pulsar.

\subsection{\label{SectChi2Surfaces}Position angle curve fitting}

The PA as a function of phase, $\psi(\phi)$, was calculated from 
the Stokes parameters for each pulsar. Any PA values for which the 
signal-to-noise ratio of the linearly polarised 
component of the emission was less than 2$\sigma$ were discarded. Only 
the remaining values were used in subsequent fitting of Eq.~\ref{EqPA}. 
In order to constrain 
the geometrical parameters, a gridsearch was performed in ($\alpha$, $\beta$) 
space. The least-$\chi^2$ fit between the RVM curve and the data was then determined by optimising the remaining free parameters $\phi_0$ and $\psi_0$. 

The obtained constraints on $\alpha$ and $\beta$ are sensitive to the 
pulse phase resolution of the data. Decreasing the resolution 
increases the signal-to-noise ratio per bin. As a 
result, the significance of the linear intensity in bins at the edges of the pulse is increased, allowing the PA to be determined over a wider range of pulse phase. However, decreasing the resolution can potentially result in
distortion of the PA swing (especially where the gradient is largest) leading to an unreliable RVM fit.
For each pulsar, $\chi^2$ surfaces were calculated using 1024, 512, 256 
and 128 bins per rotation period and the optimum resolution (see 
Table~\ref{TabDerivedParameters}) was determined and used for all 
subsequent processing of that pulsar. 

The reduced-$\chi^2$ values resulting from this fitting process 
correspond to a surface in ($\alpha$, $\beta$) space, an example of 
which can be seen in Fig.~\ref{FigJ0631+1036}. Three contours are 
shown, corresponding to 2, 3 and 4 times the global least-$\chi^2$ 
value. These represent 1$\sigma$, 2$\sigma$ and 3$\sigma$ uncertainties 
in $\alpha$ and $\beta$. The error on $\phi_0$ was 
calculated by fitting for $\alpha$, $\beta$ and $\psi_0$ repeatedly 
as $\phi_0$ was varied, and determining the value of $\phi_0$ at 
which the resulting fit had a reduced-$\chi^2$ four times that 
of the global least-$\chi^2$ fit (equivalent to a 3$\sigma$ 
error)\footnote{Note that this is equivalent to scaling 
the size of the errorbars such that the lowest reduced-$\chi^2 = 1$, 
thereby recognising the fact that there are unmodeled features in the 
observed shape of the PA curve}. 
This range and the most likely value are displayed by the horizontal 
bar above or below the PA curve in 
Figs.~\ref{FigJ0631+1036} - \ref{FigJ1835-1106}.

\subsection{\label{SectEmissionHeightConstraint}Constraint from the emission height}

Previous authors who have used RVM fitting and the A/R effect
to determine viewing geometries (e.g., \citealt{waa+10}) have 
traditionally assumed that the emission 
region fills the open field lines and hence that $\phi_{\mathrm{fid}}$
must be at the centre of the observed on-pulse region. However, the emission beam may not be symmetrically illuminated. To account for such an effect, 
the assumption that $\phi_{\mathrm{fid}}$ is at the centre of the profile 
was relaxed in this investigation. Instead a range of possible values
was determined based on the component positions, using the core-cone model as a basis \citep{ran93, kj06} and allowing for some components to be missing from the observed profile. The horizontal bar above the profile in Figs.~\ref{FigJ0631+1036} - \ref{FigJ1835-1106} displays the chosen range and our preferred value. 

Alternatively, it is possible that the asymmetry is the consequence of a beam which is populated by randomly distributed `patches' \citep{lm88}. In this case the core-cone model will not be valid. However, as the $\phi_{\mathrm{fid}}$ ranges chosen typically cover a large proportion of the pulse they are likely to be consistent with a patchy beam, and if the patches are truly random we do not expect patchy beams to cause any bias in the derived viewing geometries. This would therefore imply, for instance, that the derived $\alpha$ distribution should still describe the population as a whole.

As discussed in $\S$~\ref{SectChi2Surfaces}, a range of allowed values 
for $\phi_0$ was determined from fitting to the PA curve. 
This range, together with the range of $\phi_{\mathrm{fid}}$ determined 
from the profile shape, allows a range $\Delta\phi = \phi_0 - \phi_{fid}$ 
to be established. 
The corresponding ranges of $h_{\mathrm{em}}$ and $\rho$ were calculated using 
Eqs.~\ref{EqOffset}, \ref{EqBeamHalfOpeningAngle} and \ref{EqThetaPC}. 
Finally, if $W_{\mathrm{open}}$ is known, Eq.~\ref{EqRhoContours} provides an additional constraint. In these equations $\rho$ is taken to be the opening angle 
of a beam corresponding to an emission region confined to the 
open-field-line region. To account for the possibility of beams which are only partially illuminated we take $W_{\mathrm{open}}$ to be twice the 
separation (in phase) between the determined $\phi_{\mathrm{fid}}$ and 
the edge of the profile furthest from it. The assumption we make is therefore that at least one edge of the profile reaches the last open field line and hence, as the open-field-line region is symmetric about the fiducial plane, $W_{\mathrm{open}}$ must account for an equal amount of phase before and after $\phi_{\mathrm{fid}}$. 

There is an uncertainty in the precise location of the pulse edges due to the tapering nature of the emission and the presence of noise. The points at which the intensity was 10\% of the peak intensity were taken as the most likely to correspond to the last open field lines. This definition makes the determination of the pulse edges independent of the S/N. In most cases the uncertainties were taken to be the differences between the phases at which the intensity was 20\% and 10\% of the peak. However, in some cases such choices were clearly not representative of the pulse edge and so the limits were chosen somewhat more subjectively (see $\S$~\ref{SectResults}). The most likely location and the considered range for each pulse edge are displayed by the horizontal bars immediately below the profile in Figs.~\ref{FigJ0631+1036} - \ref{FigJ1835-1106}. As a result of specifying upper and lower limits for each edge, upper and lower values could also be determined for $W_{\mathrm{open}}$. 

The parameters $\phi_0$, $\phi_{\mathrm{fid}}$ and the pulse edges were varied within the allowed ranges 
and for each combination $\rho$ and $W_{\mathrm{open}}$ were calculated. 
Eq.~\ref{EqRhoContours} was used to derive the corresponding contour
in ($\alpha$, $\beta$) space. The full set of possible contours cover 
a region of ($\alpha$, $\beta$) space (the green regions in 
Figs.~\ref{FigJ0631+1036} - \ref{FigJ1835-1106}), which represents the possible viewing geometries consistent with the profile width and the 
offset of the PA curve with respect to the fiducial plane. Such a consideration of the uncertainties on these parameters, which allows a more complete appraisal of the possible viewing geometries for each pulsar, has not been used previously when constraining viewing geometries.
The viewing geometry should then lie within the overlap between this region and 
the 3$\sigma$ limit of the $\chi^2$ surface from RVM fitting (indicated in greyscale in Figs.~\ref{FigJ0631+1036} - \ref{FigJ1835-1106}, with a key displayed at the right-hand side of the plot).

In addition to this, a single contour (orange curve in the figures), termed the ``favoured contour'', 
was plotted for 25 of the 28 pulsars (see $\S$~\ref{SectAlphaDistribution}).  
The favoured contour was specified using the optimum $\phi_0$ value determined from the RVM fit, our preferred value for $\phi_{\mathrm{fid}}$ and the value of $W_{\mathrm{open}}$ derived using this fiducial plane position and the most likely locations of the pulse edges.

\subsection{\label{SectInterpulseConstraint}Constraint from unobserved interpulses}

In some cases a further constraint can be added to the viewing geometry 
due to the absence of an interpulse. Assuming the presence of an illuminated 
beam with half-opening angle $\rho$ at each magnetic pole, we expect 
to observe an interpulse when the line of sight passes within $\rho$ 
of both ends of the magnetic axis. This occurs when the conditions 
$\alpha - \rho < \zeta < \alpha + \rho$ and 
$180\degree - \alpha - \rho < \zeta < 180\degree - \alpha + \rho$ 
are both satisfied, where $\zeta = \alpha + \beta$ is the angle between 
the line of sight and the rotation axis. The lack of an interpulse 
indicates that the 
second condition, which can be written as 

\begin{equation}
\label{EqMissingInterpulse} -2\alpha + 180\degree - \rho < \beta < -2\alpha + 180\degree + \rho   ,
\end{equation}
is not satisfied. If the minimum allowed $\rho$ value calculated from Eq.~\ref{EqBeamHalfOpeningAngle} is non-zero, the corresponding region of 
($\alpha$, $\beta$) space can be excluded, under the given assumptions. The boundaries of this region are displayed in Figs.~\ref{FigJ0631+1036} - \ref{FigJ1835-1106} (where applicable) as red diagonal lines.

\section{Results on individual pulsars}
\label{SectResults}
 
In this section we present and discuss the constraints to the viewing geometry for each pulsar in the sample. Table~\ref{TabDerivedParameters} summarises the wavelength, rotation measure and pulse phase resolution used, the favoured values and allowed ranges determined for $\phi_0$ and $\phi_{\mathrm{fid}}$, and the subsequently calculated values of $\Delta\phi$, $W_{\mathrm{open}}$ and $h_{\mathrm{em}}$ for each pulsar. Derived half-opening angles of the beam, $\rho$, and the allowed ranges of $\alpha$ and $\beta$ are presented in Table~\ref{TabUncorrectedViewingGeometries}.

\begin{figure} 
\centering 
\includegraphics[height=0.93\hsize,angle=270]{J0631+1036_paswing.ps} 
\includegraphics[height=\hsize,angle=270]{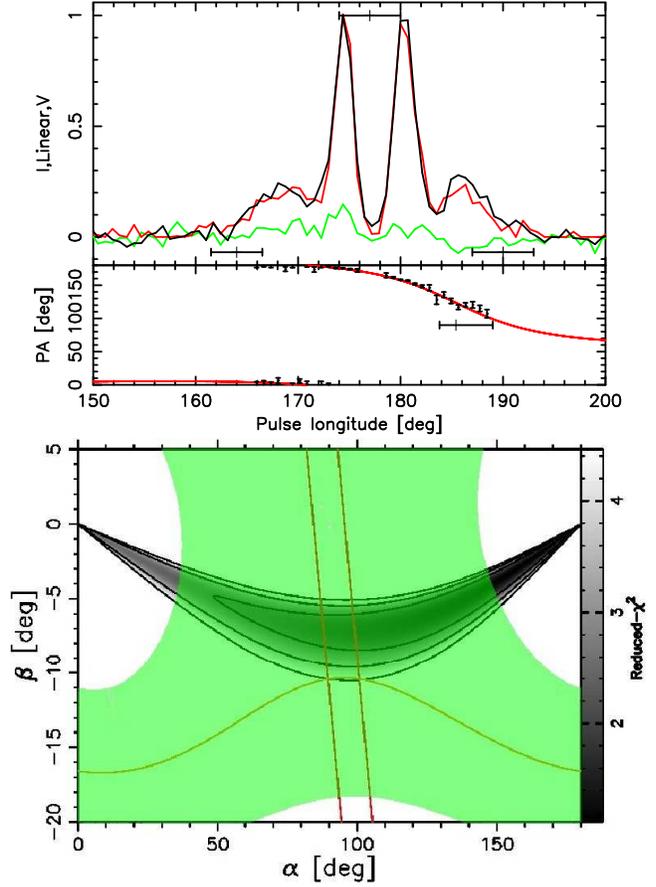} 
\caption{\label{FigJ0631+1036} PSR J0631$+$1036 at 20 cm. The upper panel in the top plot shows the profile. The black profile represents the total intensity and the red and green profiles give the intensity due to the linearly and circularly polarised components of the emission respectively. The convention is used that circular polarisation intensity is positive for left-handed polarisation and negative for right-handed polarisation. The horizontal bars show the preferred position and range of positions which were considered for the fiducial plane (above the profile) and the start and end of the pulse (below the profile). The lower panel shows the PA points with a significance greater than 2$\sigma$ superimposed with the global least-$\chi^2$ RVM fit. The horizontal bar in this panel shows the position of the inflection point including its 3$\sigma$ error. The bottom plot shows the (reduced) $\chi^2$ surface (greyscale, with contours corresponding to 1$\sigma$, 2$\sigma$ and 3$\sigma$) overlain with the solutions consistent, within the uncertainties, with the A/R effect in combination with the observed pulse width (green region). The key at the right-hand side of the plot describes the magnitude of the reduced-$\chi^2$. The `favoured' contour, which corresponds to the preferred values of $\phi_0$, $\phi_{\mathrm{fid}}$ and the pulse edges, is shown in orange. Viewing geometries between the two red (straight diagonal) lines are inconsistent with the lack of an observed interpulse. } 
\end{figure}

\subsection{PSR J0631$+$1036 (Fig.~\ref{FigJ0631+1036}) } 
 
The profile for this pulsar is highly symmetrical, which makes it likely that the fiducial plane is close to the centre of the profile. However, to ensure that the constraint on the viewing geometry is conservative we have allowed the fiducial plane to be located at any point between the two central peaks.

The constraints on the viewing geometry of this pulsar are consistent with the earlier results of \cite{waa+10}, who found $h_{em} = 600$ km and $\rho = 18\degree$. Solutions with $\sim 90\degree < \alpha <$ $\sim 100\degree$ are excluded by the lack of a visible interpulse.

\begin{figure} 
\centering 
\includegraphics[height=0.93\hsize,angle=270]{J0659+1414_paswing.ps} 
\includegraphics[height=\hsize,angle=270]{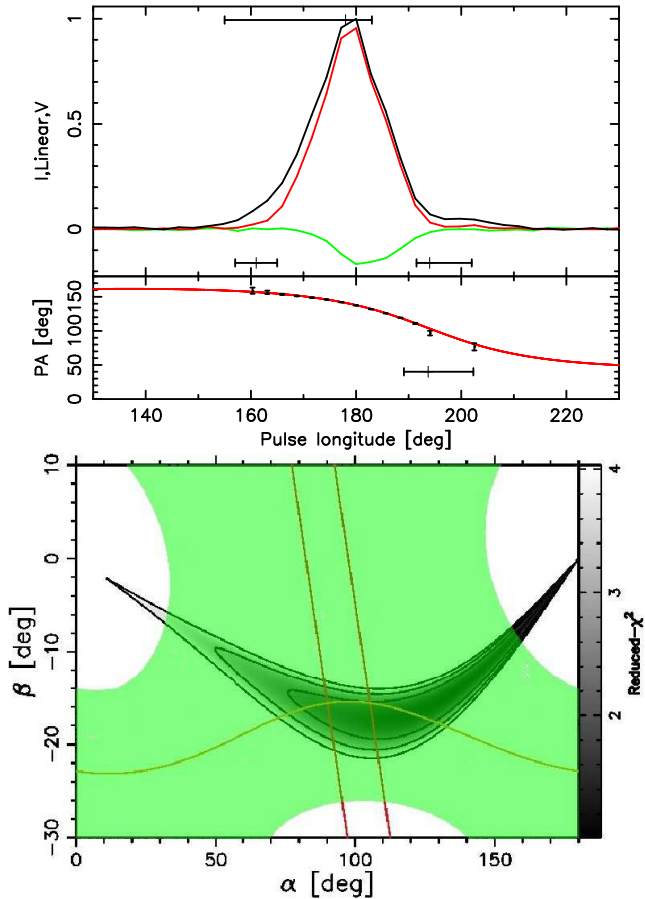} 
\caption{\label{FigJ0659+1414} PSR J0659$+$1414 at 20 cm. As Fig.~\ref{FigJ0631+1036}. } 
\end{figure}

\subsection{PSR J0659$+$1414 / B0656$+$14 (Fig.~\ref{FigJ0659+1414})  } 
 
This pulsar has a long history of polarization studies over a wide range of 
frequencies \citep{lm88, ran93, ew01, jkw06, jkk+07, waa+10, wck+04}. 
The profile is roughly triangular at our observing frequency. 
However, the profile is not symmetric about the peak. The leading edge shows depolarisation (which may be indicative of a conal component). The observed component could be the trailing component in a double, as noted for several pulsars by JW06. Therefore the possibility that the fiducial plane is before the peak was included in the allowed range. The range of possible locations of the trailing edge was extended to include the trailing component. 

Our results are on the viewing geometry are consistent with the values derived in \citep{waa+10,jw06}.

\begin{figure} 
\centering 
\includegraphics[height=0.93\hsize,angle=270]{J0729-1448_paswing.ps} 
\includegraphics[height=\hsize,angle=270]{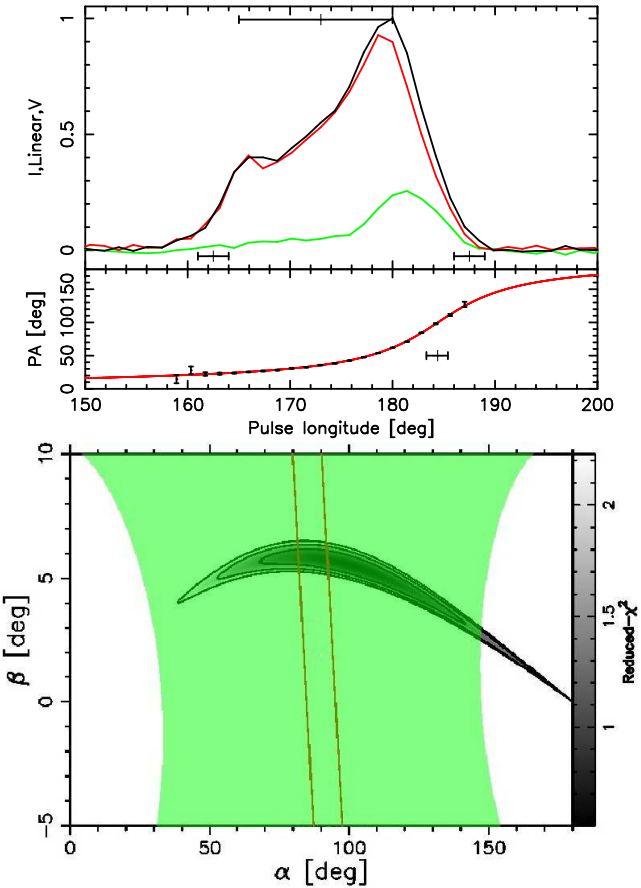} 
\caption{\label{FigJ0729-1448} PSR J0729--1448 at 20 cm. As Fig.~\ref{FigJ0631+1036}. } 
\end{figure}

\subsection{PSR J0729--1448 (Fig.~\ref{FigJ0729-1448})  } 
 
This profile is asymmetric at 1.4~GHz, which means there is ambiguity in the position of the fiducial plane. At 3.1~GHz the profile becomes double peaked, with the newly-apparent peak coinciding with the earlier component in this profile (JW06). This implies that the fiducial plane is between these two components. This is reflected in the choice of upper and lower limits on $\phi_{\mathrm{fid}}$, which correspond to the peaks of the trailing and leading components.  
 
JW06 found $h_{em} = 630$ km and a corresponding $\rho = 20\degree$. They also found $\alpha$ to be unconstrained while $0\degree < \beta < 9\degree$. Our values are consistent with these results. We can exclude $\alpha < 32\degree$ and $\alpha > 148\degree$ which in turn indicates $2\degree < \beta < 7\degree$. 
 
\begin{figure} 
\centering 
\includegraphics[height=0.93\hsize,angle=270]{J0742-2822_paswing.ps} 
\includegraphics[height=\hsize,angle=270]{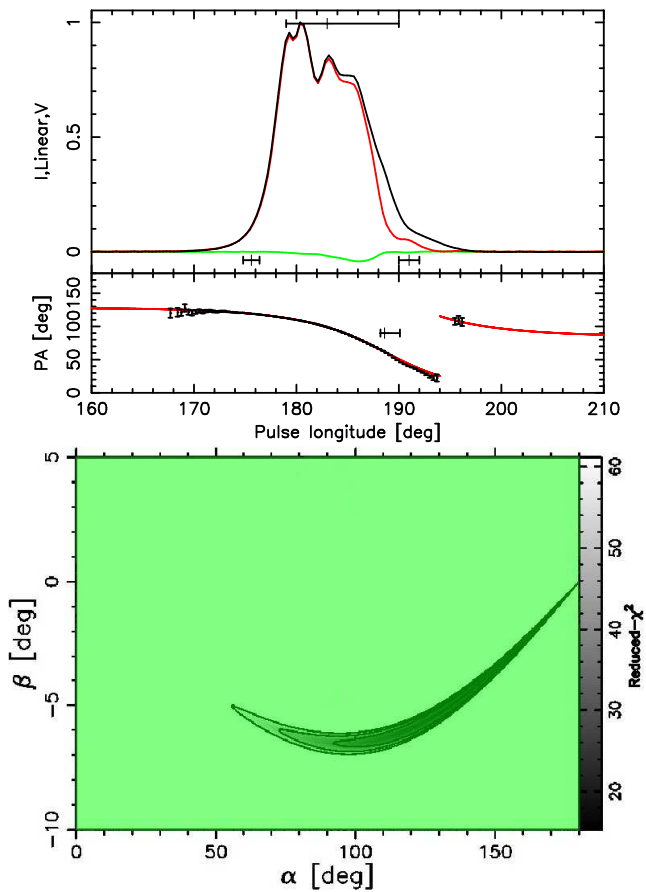} 
\caption{\label{FigJ0742-2822} PSR J0742--2822 at 20 cm. As Fig.~\ref{FigJ0631+1036}. } 
\end{figure}

\subsection{\label{J0742}PSR J0742--2822 / B0740--28 (Fig.~\ref{FigJ0742-2822})} 
 
Polarization profiles of this pulsar have been presented most recently 
in \cite{jhv+05}, \cite{kj06} and \cite{waa+10} and the pulsar was shown 
to alternate between two profile states by \cite{ksj13}. The profile 
is complex, with a boxy structure showing several prominent peaks and 
a smaller trailing component. The evolution of the profile with frequency 
is also complex \citep{jkw06}. For this reason we assign a wide range 
to $\phi_{\mathrm{fid}}$ from $179\degree$ to $190\degree$. As a consequence this does not lead to a useful additional constraint in ($\alpha$, $\beta$) space.
 
We note in passing that the orthogonal polarization mode (OPM) jump at 
the trailing edge of the profile at 3.1 GHz seen by \cite{kj06} is clearly 
seen here for the first time at 1.4 GHz, at $194\degree$ pulse phase, and 
has been included in the fit.  
 
\begin{figure} 
\centering 
\includegraphics[height=0.93\hsize,angle=270]{J0835-4510_paswing.ps} 
\includegraphics[height=\hsize,angle=270]{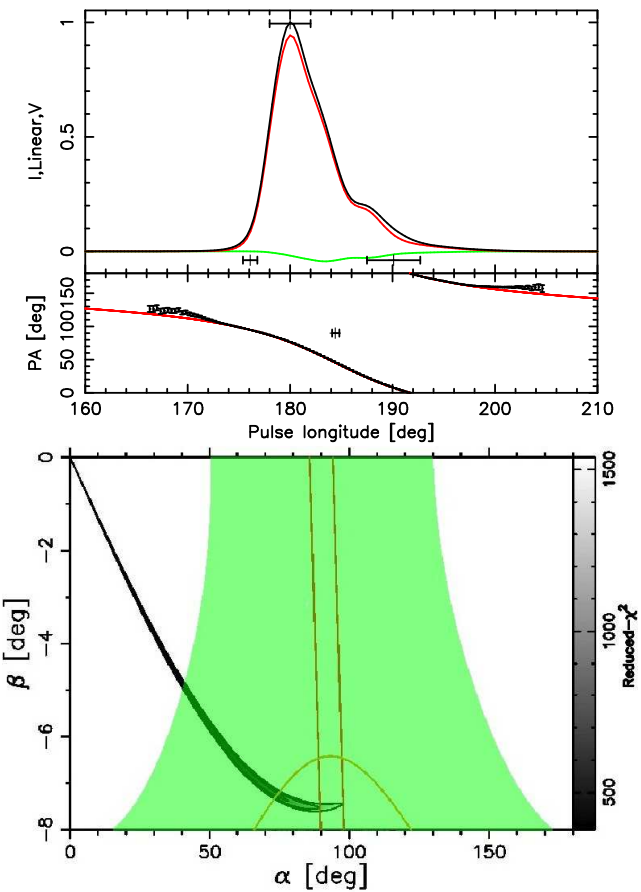} 
\caption{\label{FigJ0835-4510} PSR J0835--4510 at 20 cm. As Fig.~\ref{FigJ0631+1036}. } 
\end{figure}

\subsection{PSR J0835--4510 / B0833--45 (Fig.~\ref{FigJ0835-4510})} 
 
PSR~J0835--4510 (Vela) is the pulsar on which the RVM was based 
\citep{rc69a}. It is the brightest pulsar in the sky at this wavelength 
and emission can be seen down to very faint levels (50 dB below the 
peak). Multi-frequency observations strongly suggest that the peak seen 
at 1369~MHz is the core component followed by the trailing cone with 
the leading cone absent \citep{jvk+01, jkw06, kjl+11}. This is the motivation 
for our choice of $\phi_{\mathrm{fid}}$. 
 
\cite{jvk+01} claim $\alpha = 55\degree$, $\beta = -6\degree$ whereas 
\cite{jhv+05} have $\alpha = 43\degree$, $\beta = -6.5\degree$. X-ray 
observations of the torus by \cite{nr04} allowed a determination 
of $\alpha+\beta$ to be 64$\degree$, inconsistent with the \cite{jhv+05} 
result. Ng \& Romani suggest a solution with $\alpha = 70\degree$, 
$\beta=-6\degree$. Recent $\gamma$-ray constraints \citep{rw10} also prefer 
$\alpha$ values close to 70$\degree$. We note the \cite{jhv+05} 
solutions are not consistent with our RVM fits. 
 
Our result would be strongly affected by what we assume as the overall pulse 
width. Taking the pulse edges to be at 10\% of the peak intensity (consistent 
with the rest of the paper) leads to ($\alpha$, $\beta$) = 
($74\degree$, $-7.3\degree$). This is in good agreement the 
\cite{nr04} constraint and the $\gamma$-ray models. This scenario 
is used in Table~\ref{TabDerivedParameters} and in subsequent analysis.

\begin{figure} 
\centering 
\includegraphics[height=0.93\hsize,angle=270]{J0908-4913_MP_paswing.ps} 
\includegraphics[height=0.93\hsize,angle=270]{J0908-4913_IP_paswing.ps} 
\includegraphics[height=0.97\hsize,angle=270]{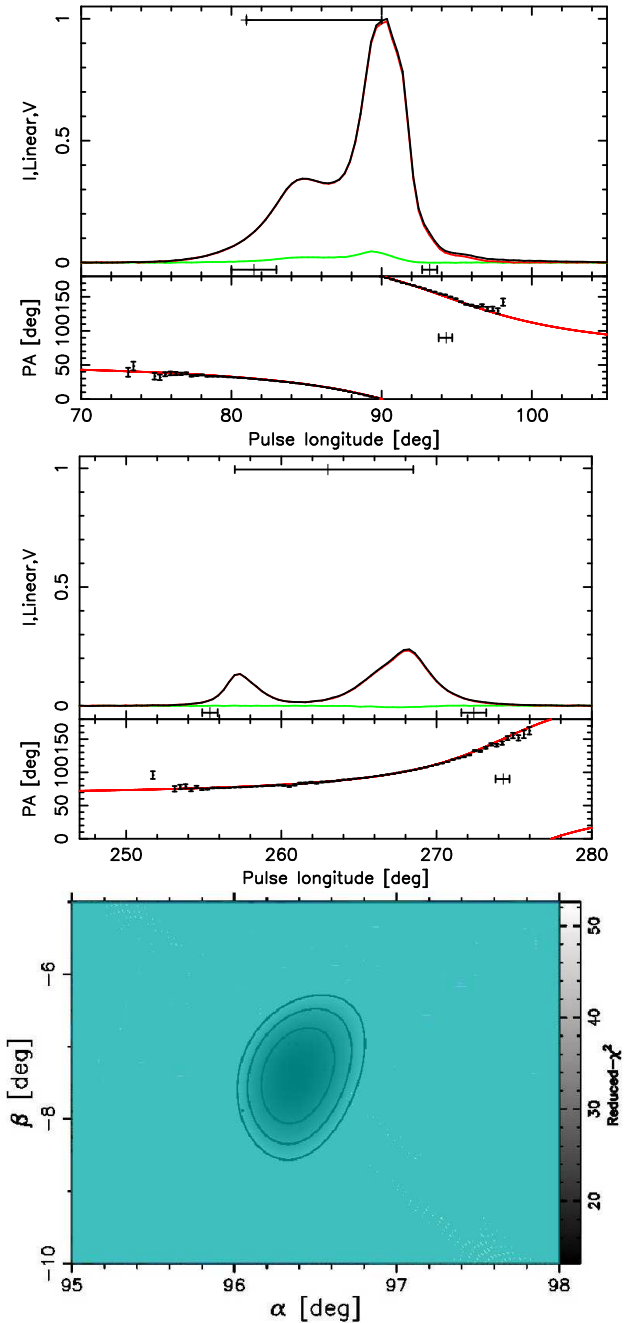} 
\caption{\label{FigJ0908-4913} PSR J0908--4913 at 20 cm. As Fig. 1. The profiles shown are the main pulse (upper panel) and the interpulse (middle panel). Both profiles are normalised to the maximum intensity of the main pulse. The edges of the interpulse were determined according to 10\% of the interpulse maximum intensity. The blue-green shading represents the superposition of the A/R constraints from the main pulse and from the interpulse. It can be seen that neither of these constraints can exclude any of the displayed region of $\alpha$ and $\beta$. } 
\end{figure}

\subsection{\label{J0908}PSR J0908--4913 / B0906--49 (Fig.~\ref{FigJ0908-4913})} 
 
This pulsar is one of two in the sample for which an interpulse is detected. \cite{kj08} found a remarkably well-constrained $\alpha \sim 96\degree$ using RVM fitting. However, they also showed that the geometry they determined varied slightly with observing frequency, from ($\alpha$, $\beta$) = (96.6$\degree$, --8.1$\degree$) at 1.4 GHz to (96.1$\degree$, --5.9$\degree$) at 8.4 GHz. They attributed this variation to a small amount of scattering by the interstellar medium. They therefore favour the viewing geometry determined at 8.4 GHz, as this higher frequency will be least affected. They suggested the fiducial plane to coincide with the centre of the interpulse and hence also with the leading component of the main pulse. This component has a high spectral index, suggesting an origin close to the magnetic axis. From this they obtain $h_{em} \sim 230$ km and $\rho \sim 18\degree$. 
 
We used a range of $\phi_{\mathrm{fid}}$ values to lie between the two components for both the main pulse and the interpulse (which is too wide to constrain the viewing geometry further). Our results agree with those of \cite{kj08} at the corresponding frequency. We therefore adopt the viewing geometry determined at 8.4 GHz by \cite{kj08} as the preferred viewing geometry. 
 
\begin{figure} 
\centering 
\includegraphics[height=0.93\hsize,angle=270]{J0940-5428_paswing.ps} 
\includegraphics[height=\hsize,angle=270]{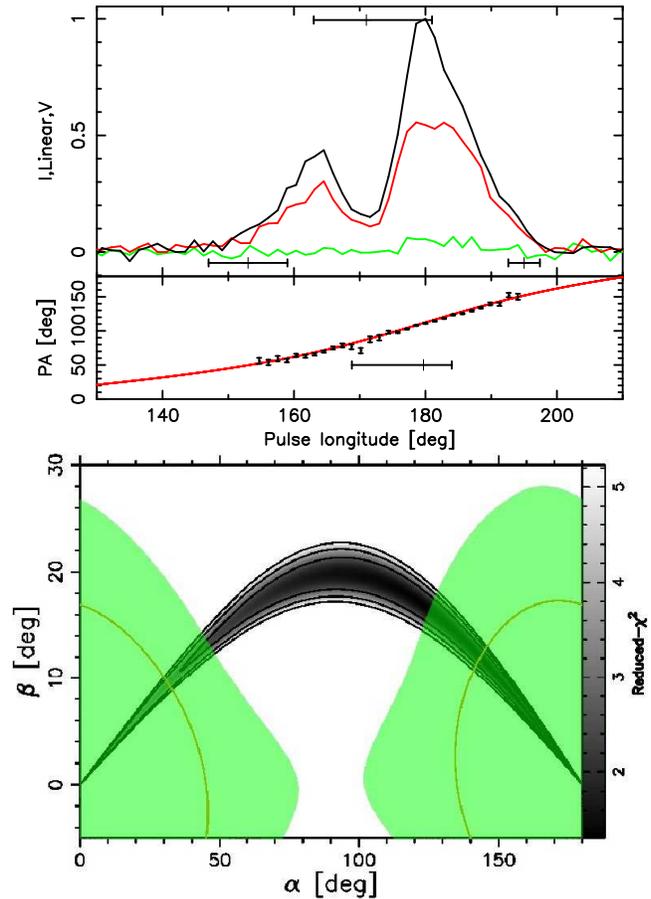} 
\caption{\label{FigJ0940-5428} PSR J0940--5428 at 20 cm. As Fig.~\ref{FigJ0631+1036}. } 
\end{figure}

\subsection{PSR J0940--5428 (Fig.~\ref{FigJ0940-5428}) } 
 
The profile for the pulsar is double peaked (JW06) and likely to be a conal double. 
The range of $\phi_{\mathrm{fid}}$ was chosen between the two components with a position approximately at the midpoint of this range being favoured.  
JW06 found $\alpha$ to be unconstrained and $\beta < 20\degree$, in agreement with our results. 
 
The profile is unusually wide, at some 60$\degree$. The relatively large proportion of the pulse period that the line of sight spends within the emission region implies a small $\alpha$ value or a large emission height. The combination of the A/R effect and the $\chi^2$ surface suggests that the emission height cannot be very large, thereby excluding $49\degree < \alpha < 122\degree$.

\begin{figure} 
\centering 
\includegraphics[height=0.93\hsize,angle=270]{J1016-5857_paswing.ps} 
\includegraphics[height=\hsize,angle=270]{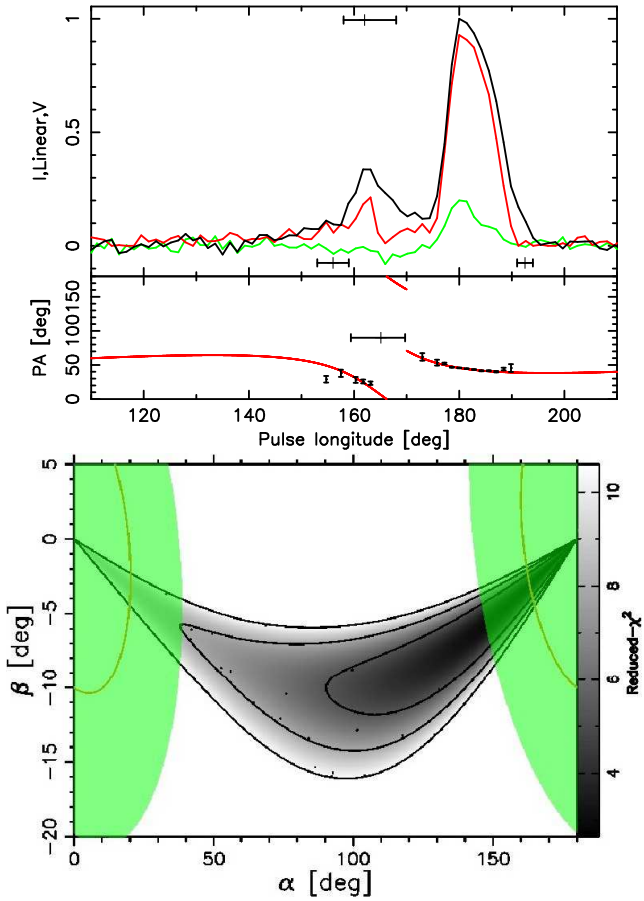} 
\caption{\label{FigJ1016-5857} PSR J1016--5857 at 20 cm. As Fig.~\ref{FigJ0631+1036}. } 
\end{figure}

\subsection{PSR J1016--5857 (Fig.~\ref{FigJ1016-5857})} 
 
This profile is a double as shown also by JW06.
The PA curve shows a large difference in PA when the two components are compared. This was found to be relatively well fitted by an OPM jump between the two components, which resulted in a $\sim 3$ times lower reduced-$\chi^2$.  
 
The inflection point occurs \emph{earlier} than the midpoint between the 
two components. This leads us to believe that we are not seeing a conal 
double, but rather a core component and a trailing cone with the leading 
cone missing. We have therefore judged $\phi_{\mathrm{fid}}$ to be located at the 
peak of the first component. The small inferred emission height implies 
that the two axes are close to alignment. This is reinforced by the large 
width of the pulse, which is $\sim 45\degree$.

\begin{figure} 
\centering 
\includegraphics[height=0.93\hsize,angle=270]{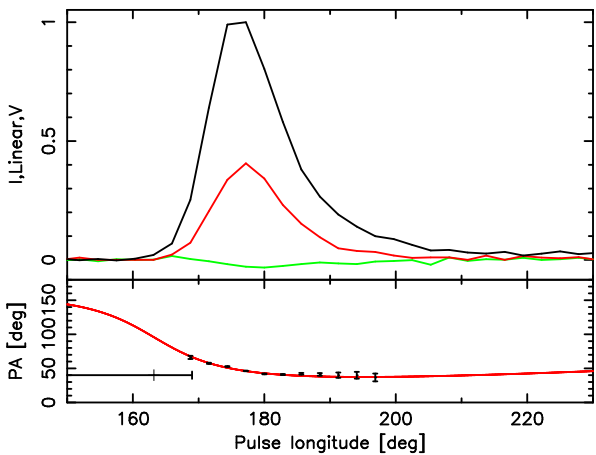} 
\caption{\label{FigJ1019-5749} PSR J1019--5749 at 10 cm. As the upper plot in Fig.~\ref{FigJ0631+1036}. } 
\end{figure}

\subsection{PSR J1019--5749 (Fig.~\ref{FigJ1019-5749})} 
 
This pulsar is highly scattered by the interstellar medium, even at 3100 MHz. Smearing of the PA curve towards later phase means the steepest part of the curve will appear to be earlier than the true position of the inflection point. Also, as the profile is distorted, it is difficult to obtain reliable estimates of the fiducial plane position and the pulse width. We therefore have not included this pulsar in further analysis. 
 
\begin{figure} 
\centering 
\includegraphics[height=0.93\hsize,angle=270]{J1028-5820_paswing.ps} 
\includegraphics[height=\hsize,angle=270]{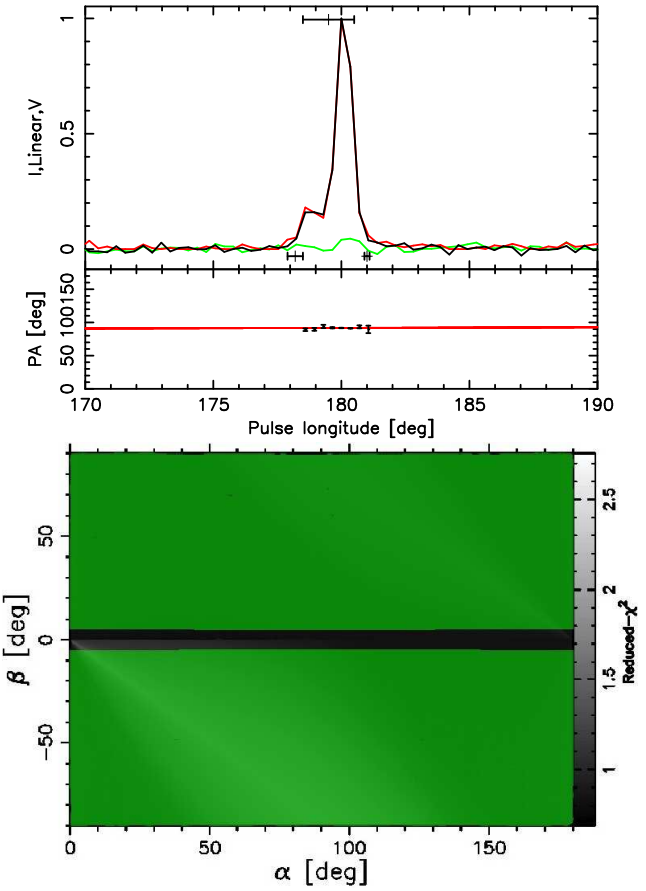} 
\caption{\label{FigJ1028-5819} PSR J1028--5819 at 20 cm. As Fig.~\ref{FigJ0631+1036}. The A/R constraint shown corresponds to the case in which the emission region fills the open-field-line region, in order to be consistent with the other plots presented in this paper. However, as discussed in the text, it appears likely that the true viewing geometry is at a smaller $\beta$ than allowed by this constraint. } 
\end{figure}

\subsection{\label{J1028}PSR J1028--5819 (Fig.~\ref{FigJ1028-5819})} 
 
This pulse profile is by far the narrowest in the sample. It was discovered 
and analysed by \cite{kjk+08}. The polarisation position angle is constant 
across the pulse, as can be seen in the figure, meaning the 
RVM curve for virtually any viewing geometry can be made to fit the data. 
There are two explanations for such a narrow profile -- a nearly diametrical 
cut across a small emission region, or a grazing cut at the edge of a larger 
emission region. The latter is unlikely to be the case for this pulsar, as 
radius-to-frequency mapping \citep{kom70} would cause the pulse width to 
be very sensitive to frequency. The pulse profiles in \cite{kjk+08} show 
no significant change in the width between 1.4 GHz and 3.1 GHz. 
 
The profile is an apparent double. The fiducial plane would then be at the 
midpoint between the two peaks. Alternatively, it is possible that either 
of these is a core component and the other a conal component, with the other 
conal component remaining undetected. For this reason the range of possible 
fiducial plane positions chosen stretches between the two peaks. 
 
The lack of a gradient of the PA curve shows that the A/R effect must be 
sufficient to move the inflection point outside the on-pulse region. This 
then gives a lower limit on $h_{\mathrm{em}}$ and hence $\rho$. We find that 
$\rho > 4\degree$ if the fiducial plane coincides with the trailing peak 
and $\rho > 9\degree$ if the fiducial plane coincides with the leading peak. 
Under the assumption that the emission region fills the open-field-line 
region, such a limit on $\rho$, coupled with the small 
value of $W_{\mathrm{open}}$, 
makes the contours virtually independent on $\alpha$ such that 
$|\beta| \approx \rho$. However, to explain the lack of frequency evolution 
of the pulse width we must assume $\beta$ to be significantly less than 
$\rho$. To allow such solutions, the illuminated part of the beam must 
be at least a factor of 2 smaller than the open-field-line region. 
 
\begin{figure} 
\centering 
\includegraphics[height=0.93\hsize,angle=270]{J1048-5832_paswing.ps} 
\includegraphics[height=\hsize,angle=270]{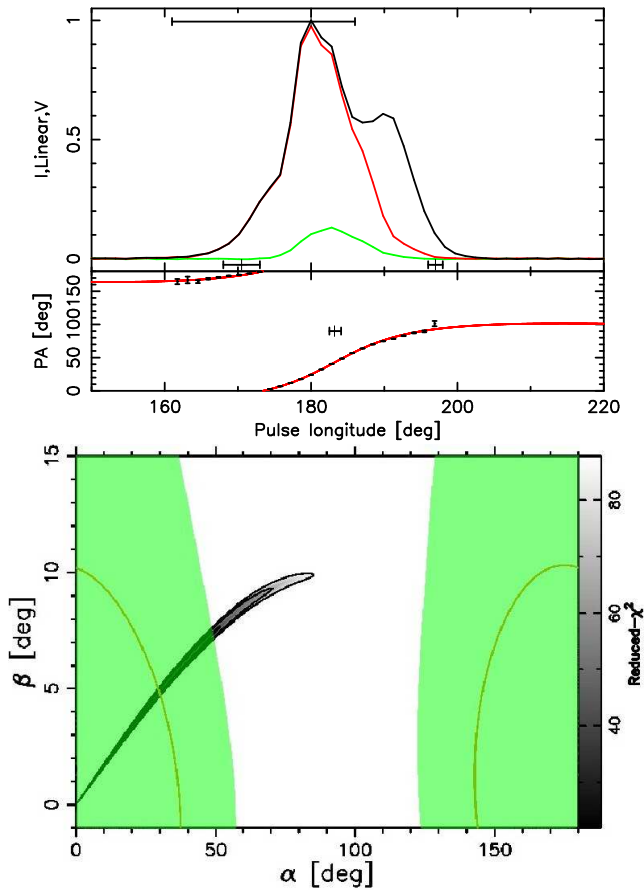} 
\caption{\label{FigJ1048--5832} PSR J1048--5832 at 20 cm. As Fig.~\ref{FigJ0631+1036}. } 
\end{figure}

\subsection{PSR J1048--5832 / B1046--58 (Fig.~\ref{FigJ1048--5832})} 
 
This pulsar is peculiar in the sample in that it shows a component with a very 
low fractional linear polarization. The profile is complex and shows a strong 
variation with frequency \citep{kjm05, jkw06}. 
 
The RVM fit for this pulsar is good and the $\chi^2$ surface alone allows values of $\alpha > 90\degree$ and values of $\beta < 0$ and $\beta > 10\degree$ to be excluded. Also, $\beta$ is exceptionally well constrained for a given $\alpha$.  
 
The position of the fiducial plane in the profile is not obvious as there is not a high degree of symmetry. At higher frequencies the profile becomes more symmetric about $\sim 180\degree$, suggesting this is the position of the fiducial plane. However, a conservative estimate $161\degree < \phi_{fid} < 186\degree$ was used, which includes all possibilities around the dominant component, and also allows for the possibility of an undetected leading component (JW06). This range allows the constraint to be refined further to $\alpha < 50\degree$, $0\degree < \beta < 7.5\degree$.  
 
\begin{figure} 
\centering 
\includegraphics[height=0.93\hsize,angle=270]{J1057-5226_MP_paswing.ps} 
\includegraphics[height=0.93\hsize,angle=270]{J1057-5226_IP_paswing.ps} 
\includegraphics[height=0.99\hsize,angle=270]{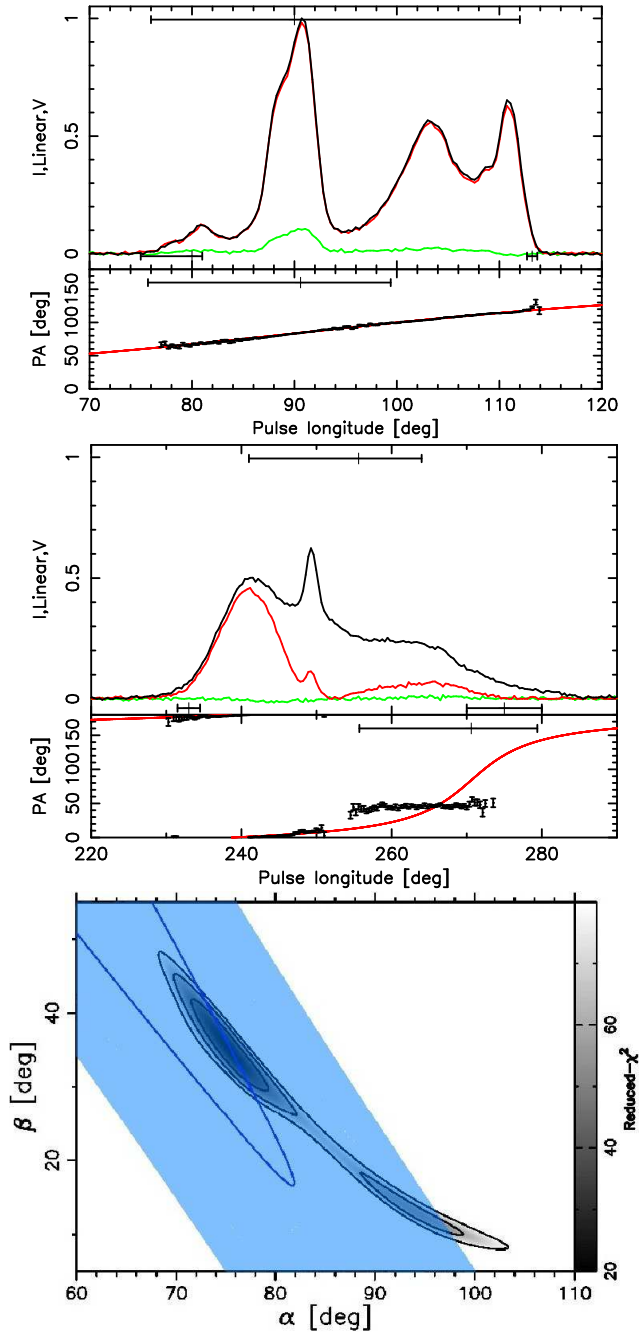} 
\caption{\label{FigJ1057-5226} PSR J1057--5226 at 20 cm. As Fig.~\ref{FigJ0631+1036}. The profiles shown are the main pulse (upper panel) and the interpulse (middle panel). Both profiles are normalised to the maximum intensity of the main pulse. The edges of the interpulse were determined according to 10\% of the interpulse maximum intensity. The constraint from the A/R effect (blue region) was derived using the interpulse only. However, $\alpha$ and $\beta$ are expressed with respect to the main pulse. } 
\end{figure}

\subsection{\label{J1057}PSR J1057--5226 / B1055--52 (Fig.~\ref{FigJ1057-5226})} 
 
This pulsar exhibits an interpulse and it has been argued that the two pulses originate from opposite poles \citep{big90,wqx+06}. An investigation into the viewing geometry by \cite{ww09} determined ($\alpha$, $\beta$) $\approx$ ($75\degree$, 36$\degree$). In that paper the authors argued that the fiducial plane was between the central spike and the trailing component of the interpulse, and hence was at the leading edge of the main pulse. Independently of this choice they concluded that the main pulse was generated outside what is conventionally thought to be the open-field-line region, with the trailing edge of the main pulse most likely originating a factor of two further from the magnetic axis than the last open field lines. 
 
The RVM fit to the PA curve is good for the main pulse and the leading half of the interpulse. However, there is an abrupt jump in $\psi$ at the approximate centre of the interpulse, which is accompanied by complete depolarisation. This deviation is more pronounced than in the data of \cite{ww09} due to the increase in S/N. It is not possible to find an RVM curve which will simultaneously fit the data on both sides of the jump. Despite this, the $\chi^2$ surface presented here is well constrained and consistent with the viewing geometry of the earlier paper. 
 
According to Weltevrede \& Wright the emission region responsible for the main pulse is not confined to the conventional open-field-line region. For this reason the constraint from the emission height and pulse width presented here was calculated using the interpulse only. Possible positions of the fiducial plane between the leading and trailing components ($241\degree < \phi_{fid} < 264\degree$) were considered, with the favoured value coinciding with the fiducial plane position used in the earlier paper. The resulting constraint (shown in blue in the figure) is consistent with the $\chi^2$ surface. The favoured contour is consistent with the viewing geometry reported by Weltevrede \& Wright.

\begin{figure} 
\centering 
\includegraphics[height=0.93\hsize,angle=270]{J1105-6107_paswing.ps} 
\includegraphics[height=\hsize,angle=270]{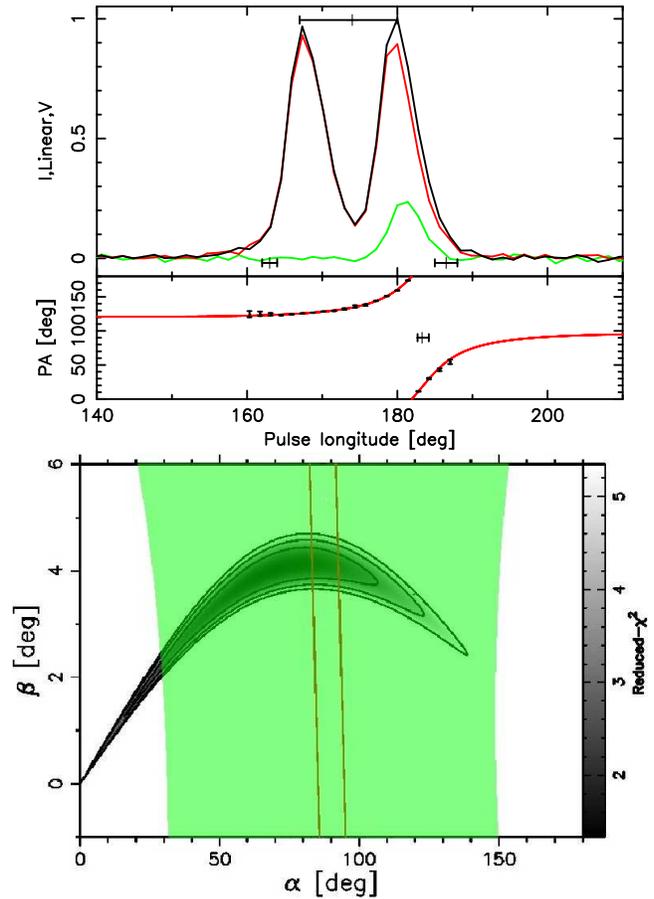} 
\caption{\label{FigJ1105--6107} PSR J1105--6107 at 20 cm. As Fig.~\ref{FigJ0631+1036}. } 
\end{figure}

\subsection{PSR J1105--6107 (Fig.~\ref{FigJ1105--6107})} 
\label{J1105}
 
JW06 found $0\degree < \beta < 4\degree$ and $\alpha$ to be unconstrained for this pulsar. Our $\chi^2$ surface is less well constrained in $\beta$, although we can exclude $\beta > 5\degree$ at the 3$\sigma$ level. We can also exclude $\alpha > 140\degree$ based on the RVM fit alone. 
 
The profile shows a high degree of symmetry at this frequency. This suggests that the fiducial plane is at the centre of the profile. However, at 3100 MHz the trailing component is significantly more intense than the leading component (JW06). This difference in spectral index might suggest that one of the components is a core component, and for this reason the fiducial plane range was taken as $167\degree < \phi_{fid} < 180\degree$. Combination of this constraint with the $\chi^2$ surface indicates that $28\degree < \alpha < 140\degree$. 
 
\begin{figure} 
\centering 
\includegraphics[height=0.93\hsize,angle=270]{J1112-6103_paswing.ps} 
\includegraphics[height=\hsize,angle=270]{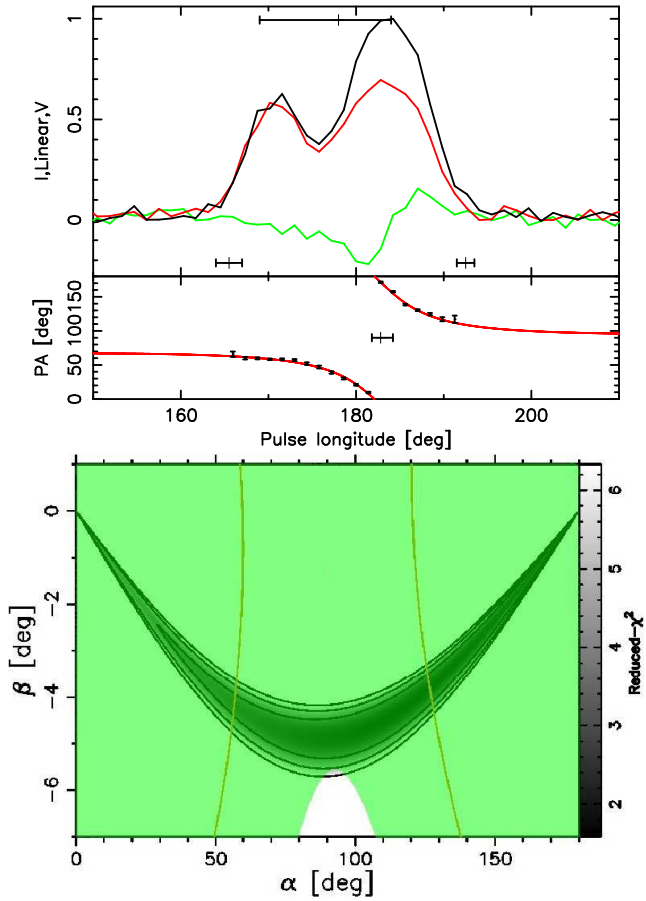} 
\caption{\label{FigJ1112-6103} PSR J1112--6103 at 10 cm. As Fig.~\ref{FigJ0631+1036}. } 
\end{figure}

\subsection{PSR J1112--6103 (Fig.~\ref{FigJ1112-6103})} 
 
The reduced-$\chi^2$ of the RVM fit is good for this pulsar. The $\chi^2$ surface is correspondingly well constrained in $\beta$, giving $-5.5\degree < \beta < 0$. The positions of the two peaks were used as the limits on the fiducial plane position. This only marginally improves the constraint in ($\alpha$, $\beta$) space.

\begin{figure} 
\centering 
\includegraphics[height=0.93\hsize,angle=270]{J1119-6127_combined_paswing.ps} 
\includegraphics[height=0.97\hsize,angle=270]{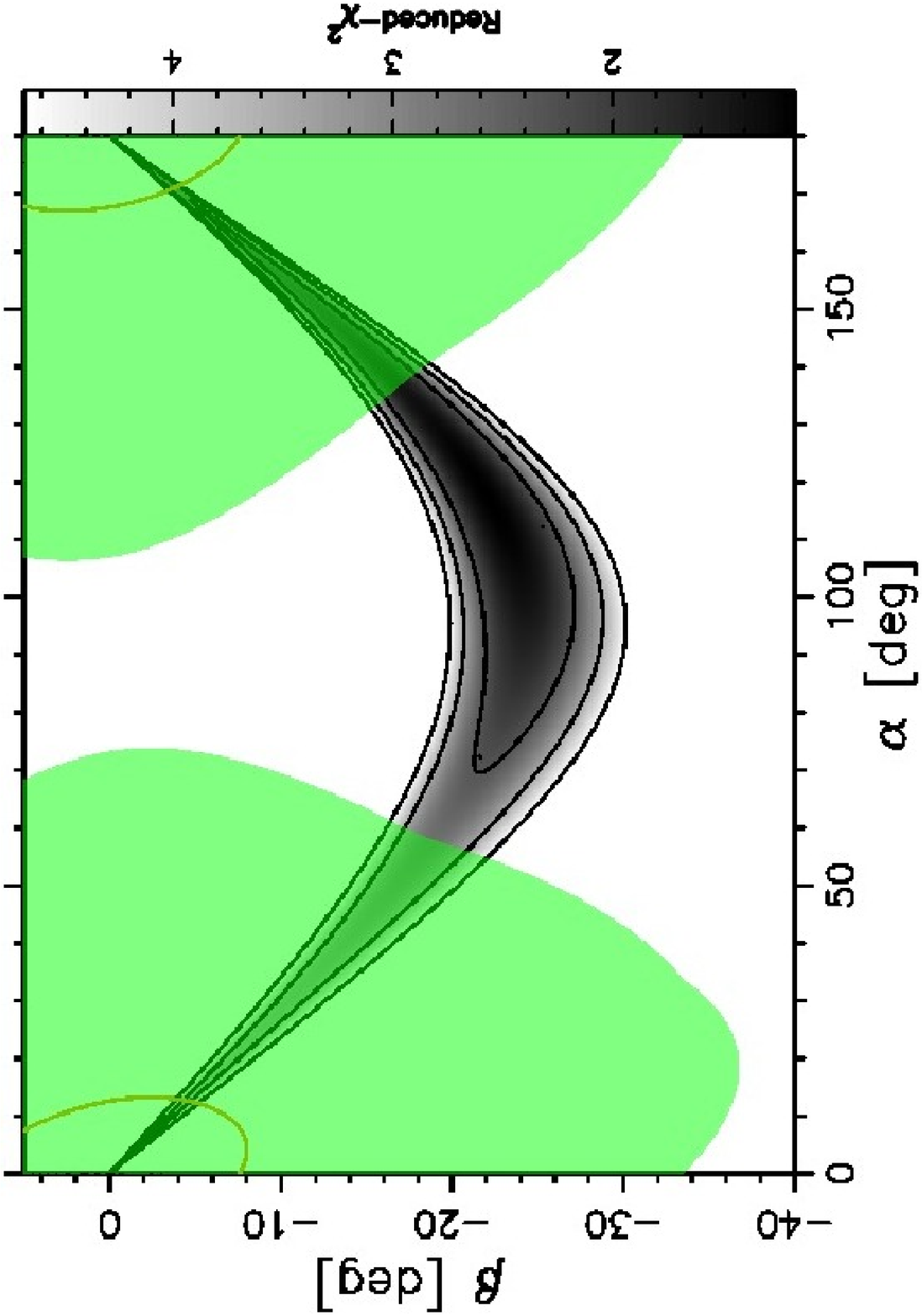} 
\includegraphics[height=0.97\hsize,angle=270]{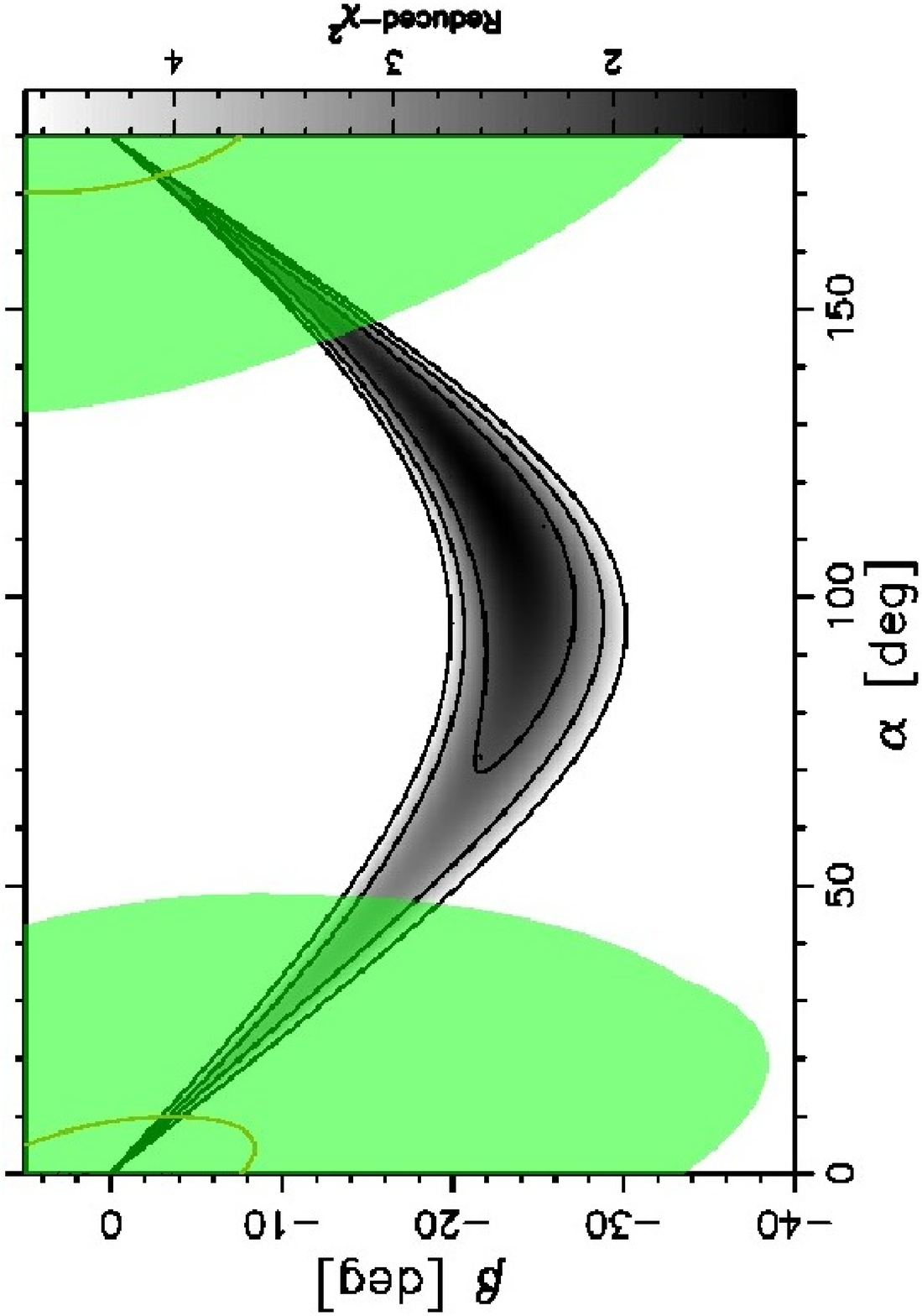} 
\caption{\label{FigJ1119-6127} PSR J1119--6127 at 20 cm. As Fig.~\ref{FigJ0631+1036}. In addition, the dashed line describes the total intensity of the double-peaked profile as reported by \protect \cite{wje11}. The PA values at $\phi > 200\degree$ (marked x) are from the double-peaked profile and fitting was performed on the concatenated PA curve. The constraints are shown for the cases with the RRAT-like components (not visible in the profile) outside (middle panel) and inside (bottom panel) the open-field-line region. } 
\end{figure}

\subsection{\label{J1119}PSR J1119--6127 (Fig.~\ref{FigJ1119-6127})} 
 
This pulsar usually exhibits a single component as shown by the solid lines in the figure and by JW06. However, on one occasion \citep{wje11} a second, comparable component was observed, the peak of which was $\sim 20\degree$ later in phase than the commonly observed peak (shown as the dashed line in the figure). In the same paper, two RRAT-like components were reported, flanking the two main components. 
 
Although the PA curve obtained from our data contains points over a smaller range of phase ($155\degree < \phi < 200\degree$) than that in \cite{wje11}, our improved S/N results in a more constrained $\chi^2$ surface from RVM fitting. We also included the values from \cite{wje11} at later phase, which allowed the constraint to be improved further.

\cite{wje11} discussed possible alignments of the double-peaked profile relative to the single-peaked profile and argued that only a coincidence between the peaks of the single profile and the leading component of the double profile is plausible. If the double-peaked profile was presumed to occur earlier relative to the single-peaked profile the values derived here for the emission height, beam half-opening angle and $\alpha$ would all increase.

We consider two situations, one in which the RRAT-like components are outside the open-field-line region and the other in which they are inside the open-field-line region. In both cases, the component configuration would have a high degree of mirror symmetry about a pulse phase $\sim 190\degree$. This therefore seems the likely position of the fiducial plane, although the range of allowed values was taken to be between the two peaks of the double profile. The range of $\rho$ values was the same for the two situations. 
 
The choice of location of the pulse edges depends on whether or not the RRAT-like emission was generated on open field lines. In the first situation (outer components outside the open-field-line region) the region $62\degree < \alpha < 132\degree$ was excluded (middle panel in the figure). In the other situation this exclusion extended to $48\degree < \alpha < 144\degree$ (bottom panel in the figure). It is unclear which situation is correct. Both scenarios are quoted in Table~\ref{TabDerivedParameters}. The favoured contour used in subsequent analysis was that corresponding to the situation in which the RRAT-like components are outside the open-field-line region, as this scenario provides the more conservative constraint on the viewing geometry. However, the favoured geometry is similar in the two situations, so this choice is not critical to the conclusions of this paper.
 
\begin{figure} 
\centering 
\includegraphics[height=0.93\hsize,angle=270]{J1357-6429_paswing.ps} 
\includegraphics[height=\hsize,angle=270]{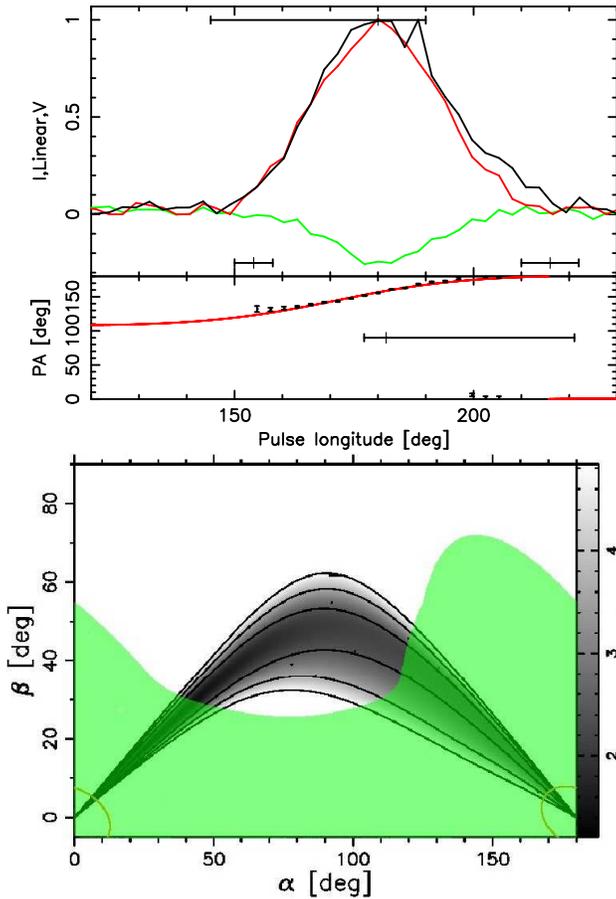} 
\caption{\label{FigJ1357-6429} PSR J1357--6429 at 20 cm. As Fig.~\ref{FigJ0631+1036}. } 
\end{figure}

\subsection{PSR J1357--6429 (Fig.~\ref{FigJ1357-6429})} 
 
The profile is symmetric, single and broad with a shallow PA curve resulting in 
a relatively poor constraint on $\beta$ from RVM fitting alone and
we find that $0\degree < \beta < 60\degree$.  This is at odds with
\cite{lzg+11} who appear however to have used a value of the slope of the PA curve which is much too high.

The symmetry in the profile suggests that the fiducial plane corresponds to the peak ($\sim 180\degree$), close to the inflection point of the PA curve. This means that $\rho$ is likely to be small (although the large error on $\phi_0$ allows significantly larger $\rho$ values). Also, the profile is wide. These two effects suggest that the axes are close to being aligned. To allow a scenario in which the leading peak of a double-peaked profile is unobserved (JW06), a conservative fiducial plane range was chosen. The combination of the two constraints shows that $\alpha < 55\degree$ or $> 102\degree$ and that $0\degree < \beta < 50\degree$, while the favoured contour suggests that both $\alpha$ and $\beta$ are significantly smaller than this. 
 
\begin{figure} 
\centering 
\includegraphics[height=0.93\hsize,angle=270]{J1410-6132_paswing.ps} 
\includegraphics[height=\hsize,angle=270]{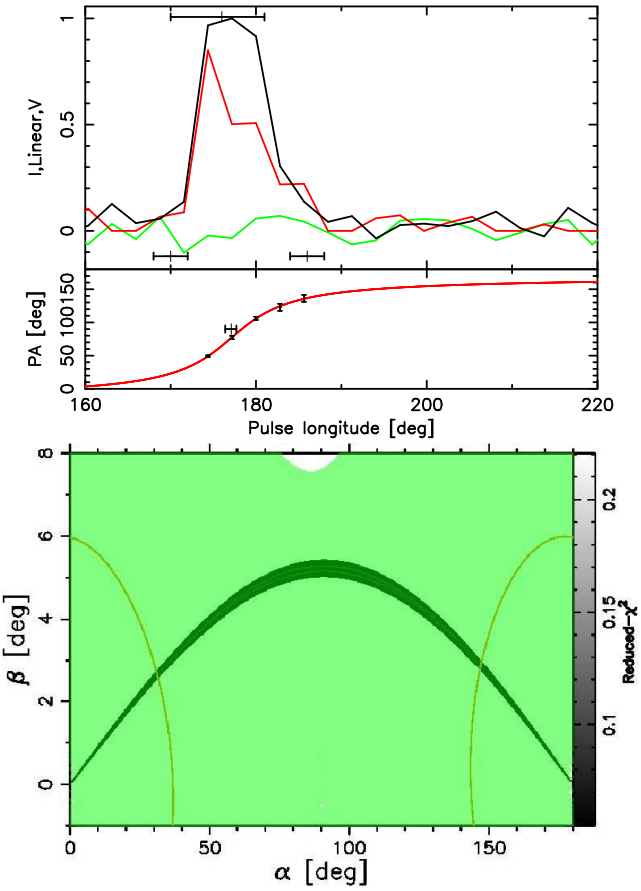} 
\caption{\label{FigJ1410-6132} PSR J1410--6132 at 5 cm. As Fig.~\ref{FigJ0631+1036}. } 
\end{figure}

\subsection{\label{J1410}PSR J1410--6132 (Fig.~\ref{FigJ1410-6132})} 
 
The profile is highly scattered at 20 cm and at 10 cm. For this reason we used archival data taken at 5 cm using the Parkes telescope in 2007 November. The profile exhibits a single component. A conservative estimate of the fiducial plane position was adopted, which includes virtually the entire pulse.  
 
The PA curve is well fitted by the RVM with only a small error on $\phi_0$.  The resulting $\chi^2$ surface shows that $0\degree < \beta < 5.5\degree$, with an excellent constraint on $\beta$ for a given $\alpha$. The conservative $\phi_{\mathrm{fid}}$ estimate means that the A/R effect cannot constrain the geometry further. 
 
\begin{figure} 
\centering 
\includegraphics[height=0.93\hsize,angle=270]{J1420-6048_paswing.ps} 
\includegraphics[height=\hsize,angle=270]{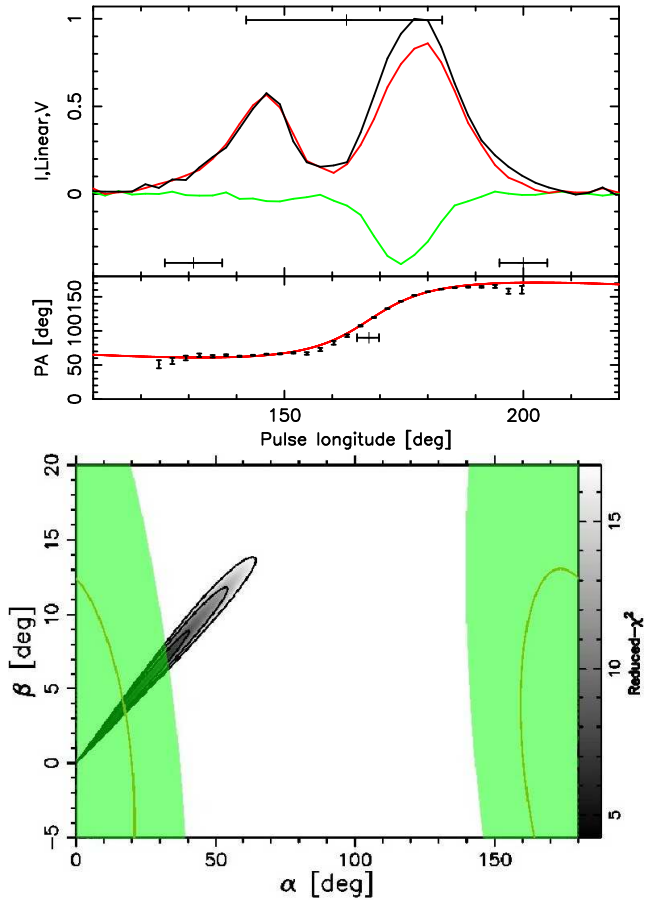} 
\caption{\label{FigJ1420-6048} PSR J1420--6048 at 20 cm. As Fig.~\ref{FigJ0631+1036}. } 
\end{figure}

\subsection{PSR J1420--6048 (Fig.~\ref{FigJ1420-6048})} 
 
The profile of this pulsar is wide and double (see also \citealt{jw06, waa+10}).

The higher S/N of our data allows an improved RVM fit. The $\chi^2$ surface in the figure shows that $\alpha < 70\degree$ and $\beta < 15\degree$. Although more constrained, the surface is consistent with that of \cite{waa+10}. Assuming the fiducial plane is between the two peaks allows the viewing geometry to be constrained to $\alpha < 33\degree$, $0\degree < \beta < 8.5\degree$ when aberration is taken into account.

\begin{figure} 
\centering 
\includegraphics[height=0.93\hsize,angle=270]{J1509-5850_paswing.ps} 
\includegraphics[height=\hsize,angle=270]{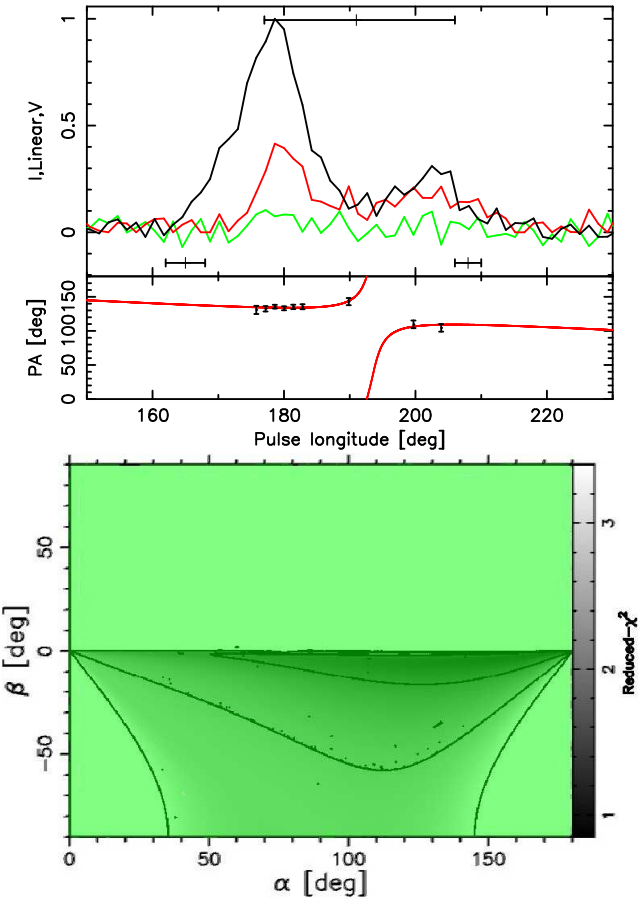} 
\caption{\label{FigJ1509-5850} PSR J1509--5850 at 20 cm. As Fig.~\ref{FigJ0631+1036}. No constraint from the emission height and pulse width is possible as the nature of the best fit, and therefore $\phi_0$, is highly uncertain.} 
\end{figure}

\subsection{PSR J1509--5850 (Fig.~\ref{FigJ1509-5850})} 
 
The S/N of the profile is low but shows a double component similar to 
that shown in \cite{waa+10}. The percentage linear polarisation was found in that paper to be relatively low for this pulsar and the data prevented the authors constructing a PA curve or an RVM fit. 

We find a similarly low percentage linear polarisation, but the increased S/N of our data allows a PA curve to be measured, a first for this pulsar. Similarly to PSR J1028--5819 ($\S$~\ref{J1028}), the limited range over which the PA can be determined, coupled with its shallow gradient, means that the RVM curve for virtually any viewing geometry can be made to fit the data. We therefore exclude this pulsar from further analysis.
 
\begin{figure} 
\centering 
\includegraphics[height=0.93\hsize,angle=270]{J1513-5908_paswing.ps} 
\includegraphics[height=\hsize,angle=270]{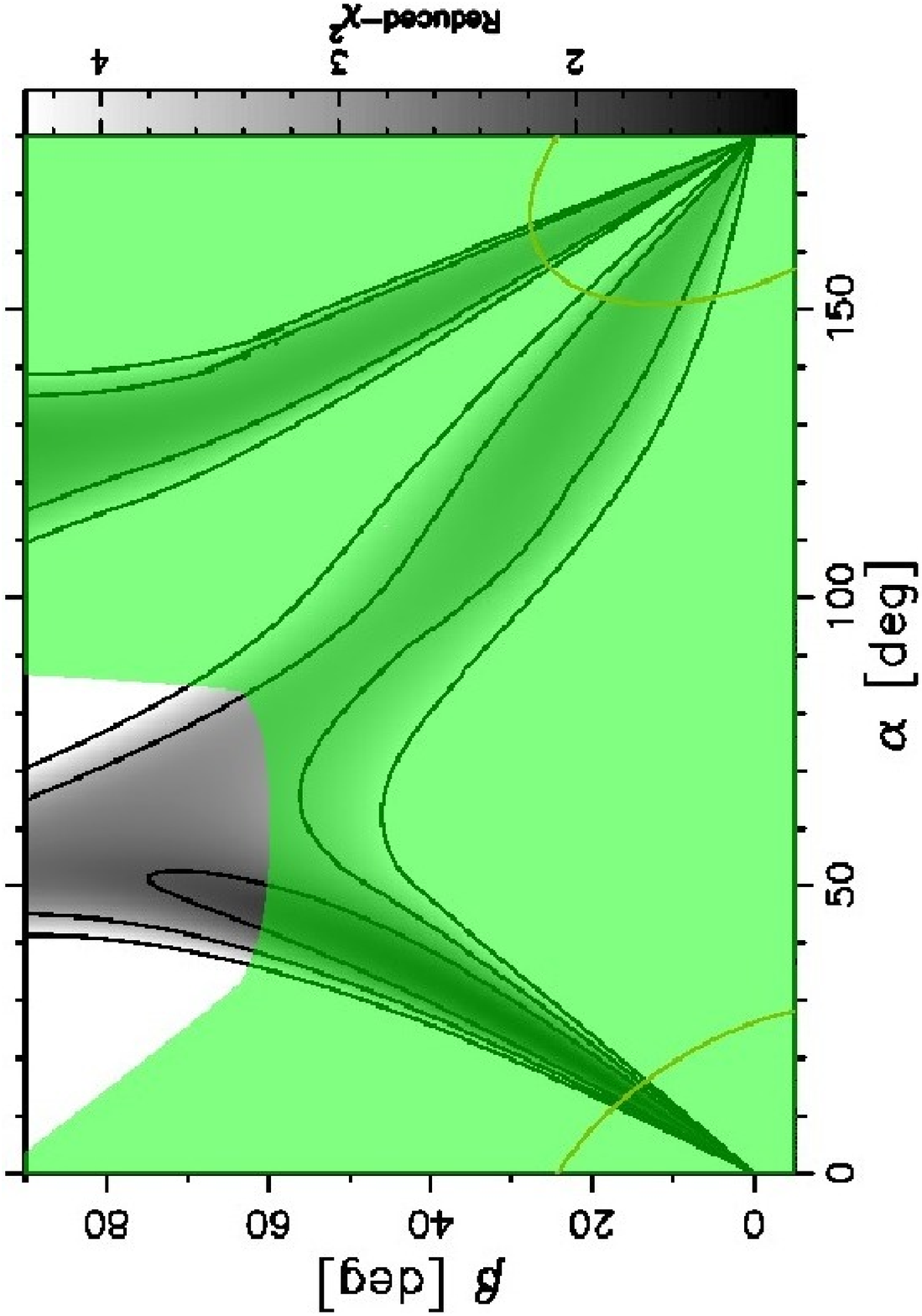} 
\caption{\label{FigJ1513-5908} PSR J1513--5908 at 20 cm. As Fig.~\ref{FigJ0631+1036}. } 
\end{figure}

\subsection{PSR J1513--5908 / B1509--58 (Fig.~\ref{FigJ1513-5908})} 
 
The profile appears as a single component with a very low intensity leading shoulder at pulse phase $120\degree - 145\degree$. This leading shoulder appears more significant at higher frequencies (JW06). The larger degree of circular polarisation in the trailing component is a feature typical of double profiles in young pulsars (e.g., JW06). Hence, the fiducial plane position range was chosen as $120\degree < \phi_{fid} < 187\degree$. However, to allow for the possibility that the leading shoulder does not originate within the open-field-line region, the allowed range for the location of the leading edge was extended to the leading edge of the trailing component.  
The measured PA curve is very well fitted by the RVM. However, as the curve is relatively shallow, neither $\alpha$ nor $\beta$ can be constrained from the $\chi^2$ surface alone.  
 
The 3$\sigma$ error on $\phi_0$ is larger than the range of phase shown in the figure, indicating that the emission height, and hence $\rho$, are highly uncertain. As a result, the combined constraint from the $\chi^2$ surface and the A/R effect is unable to constrain $\alpha$ and provides a relatively poor $\beta$ constraint, $\beta < 70\degree$.   
 
\begin{figure} 
\centering 
\includegraphics[height=0.93\hsize,angle=270]{J1531-5610_paswing.ps} 
\includegraphics[height=\hsize,angle=270]{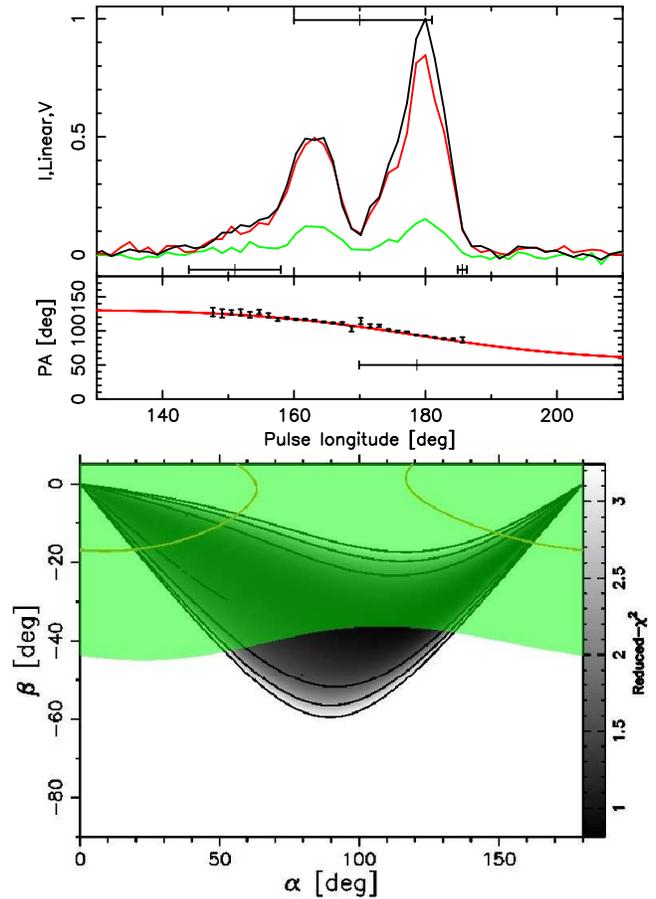} 
\caption{\label{FigJ1531-5610} PSR J1531--5610 at 20 cm. As Fig.~\ref{FigJ0631+1036}. } 
\end{figure}

\subsection{PSR J1531--5610 (Fig.~\ref{FigJ1531-5610})} 
 
The PA curve is shallow, meaning that $\beta$ is relatively poorly constrained from the $\chi^2$ surface. The profile exhibits two components, so the limits on the position of the fiducial plane were chosen to coincide with the two peaks and the midpoint between the two peaks was chosen as the favoured value of $\phi_{\mathrm{fid}}$. The resulting constraint suggests that $-43\degree < \beta < 0$ but leaves $\alpha$ unconstrained.

\begin{figure} 
\centering 
\includegraphics[height=0.93\hsize,angle=270]{J1648-4611_paswing.ps} 
\includegraphics[height=\hsize,angle=270]{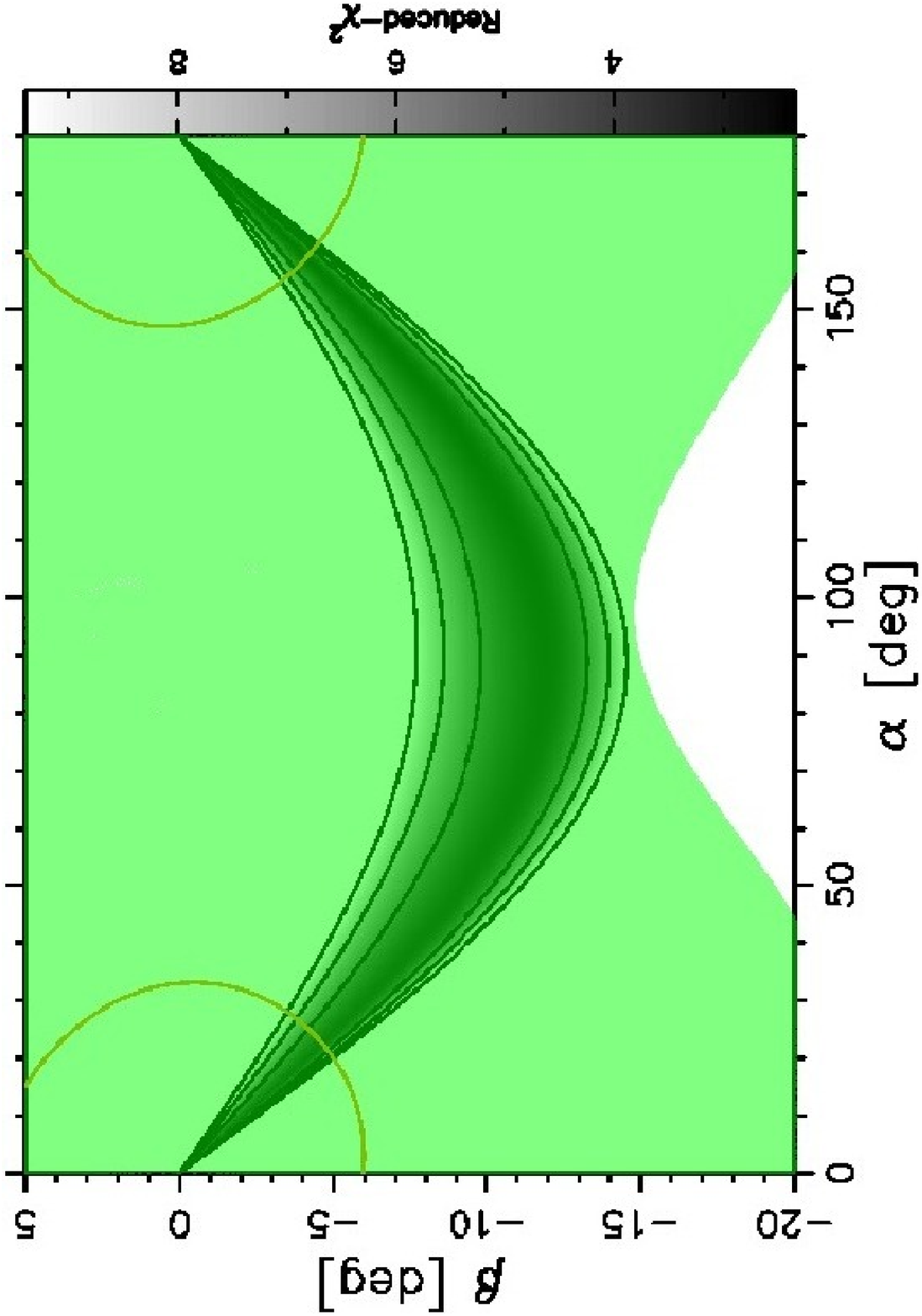} 
\caption{\label{FigJ1648-4611} PSR J1648--4611 at 10 cm. As Fig.~\ref{FigJ0631+1036}. } 
\end{figure}

\subsection{PSR J1648--4611 (Fig.~\ref{FigJ1648-4611})} 
 
The profile is scattered at 1.4~GHz, hence we show the 3.1~GHz profile here. 
The profile is symmetric giving a strong indication that the fiducial plane is at the midpoint between the two peaks. However, to account for the possibility of a missing component either before or after the observed profile, all phases between the two peaks were included in the range of possible positions of the fiducial plane. The $\chi^2$ surface is relatively well constrained, to $-15\degree < \beta < 0$. The relatively large error on $\phi_0$ means that neither $\alpha$ nor $\beta$ can be further constrained by the A/R effect. 
 
\begin{figure} 
\centering 
\includegraphics[height=0.93\hsize,angle=270]{J1702-4128_paswing.ps} 
\includegraphics[height=\hsize,angle=270]{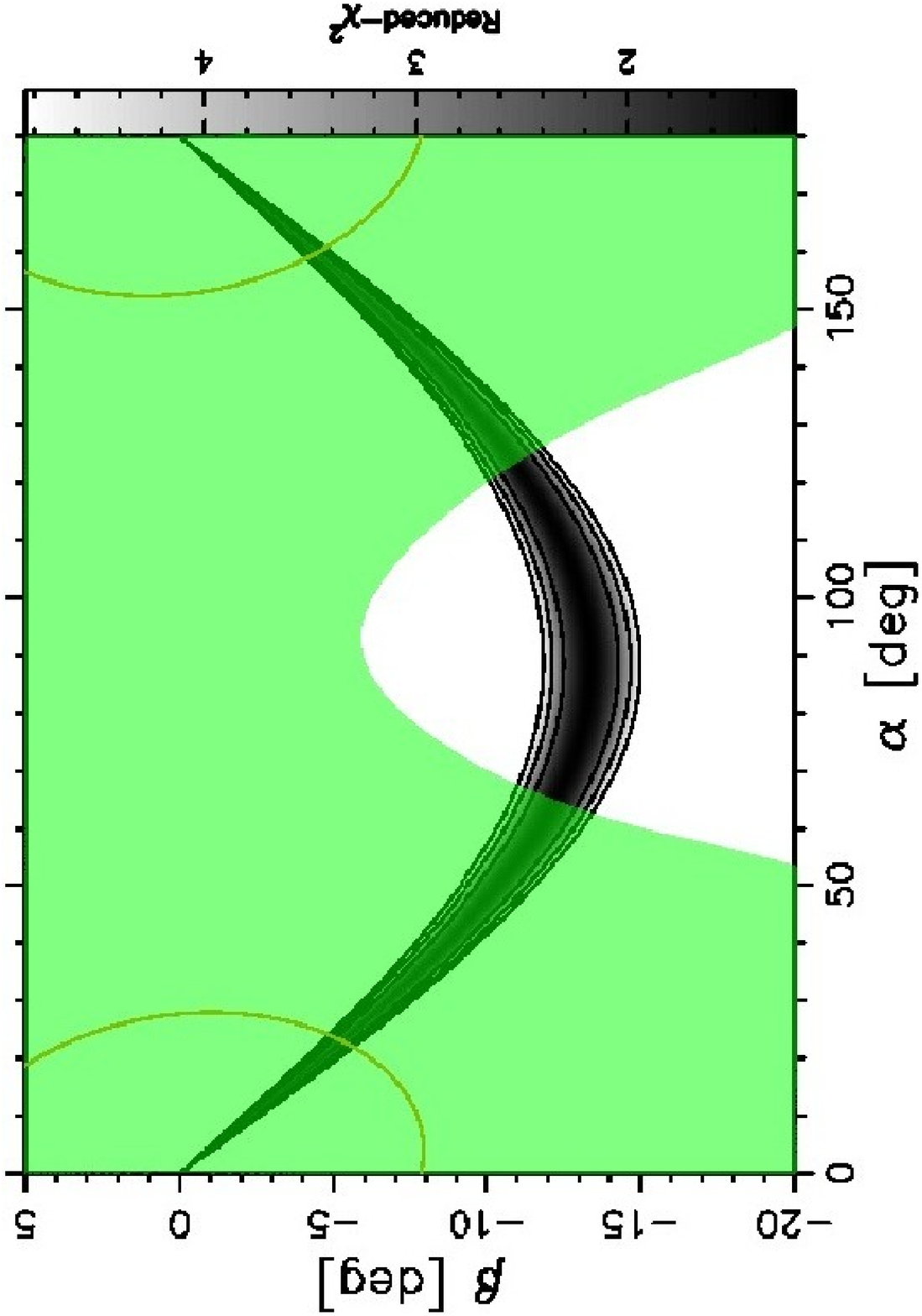} 
\caption{\label{FigJ1702-4128} PSR J1702--4128 at 10 cm. As Fig.~\ref{FigJ0631+1036}. } 
\end{figure}

\subsection{PSR J1702--4128 (Fig.~\ref{FigJ1702-4128})} 
 
The profile contains a dominant component with a shoulder at its leading edge, $\sim 8\degree$ before the peak. This suggests that the profile is an overlapping double. Judging by the profile the fiducial plane is between the two components. However, as for other pulsars in this sample we allow for a missing leading component (JW06). The combined constraint shows that solutions in the region $68\degree < \alpha < 120\degree$ can be excluded. 
 
\begin{figure} 
\centering 
\includegraphics[height=0.93\hsize,angle=270]{J1709-4429_paswing.ps} 
\includegraphics[height=\hsize,angle=270]{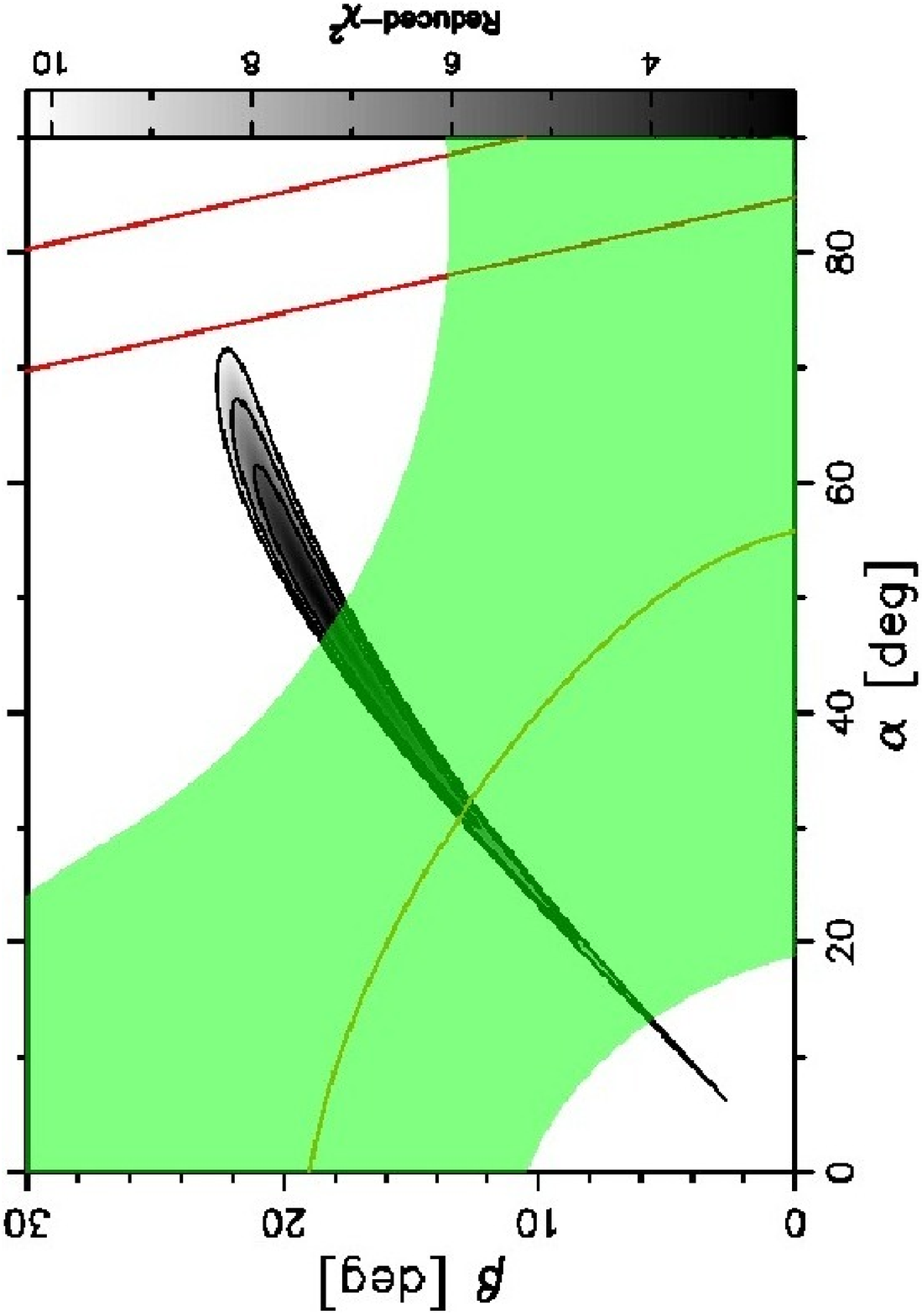} 
\caption{\label{FigJ1709-4429} PSR J1709--4429 at 20 cm. As Fig.~\ref{FigJ0631+1036}. } 
\end{figure}

\subsection{PSR J1709--4429 / B1706--44 (Fig.~\ref{FigJ1709-4429})} 
 
RVM fitting alone imposes the limit $5\degree < \alpha < 73\degree$, $2.5\degree < \beta < 24\degree$ on the viewing geometry for this pulsar. The profile is a single component with a high degree of symmetry over a wide range of radio frequencies \citep{jhv+05, kjm05, jkw06}. However, to account for the possibility of the observed component being the trailing side of a double, the chosen range of fiducial plane position was extended considerably towards earlier pulse phase. When combined with the $\chi^2$ surface this predicts $12\degree < \alpha < 50\degree$ and $5.5\degree < \beta < 19\degree$.  
 
\begin{figure} 
\centering 
\includegraphics[height=0.93\hsize,angle=270]{J1718-3825_paswing.ps} 
\includegraphics[height=\hsize,angle=270]{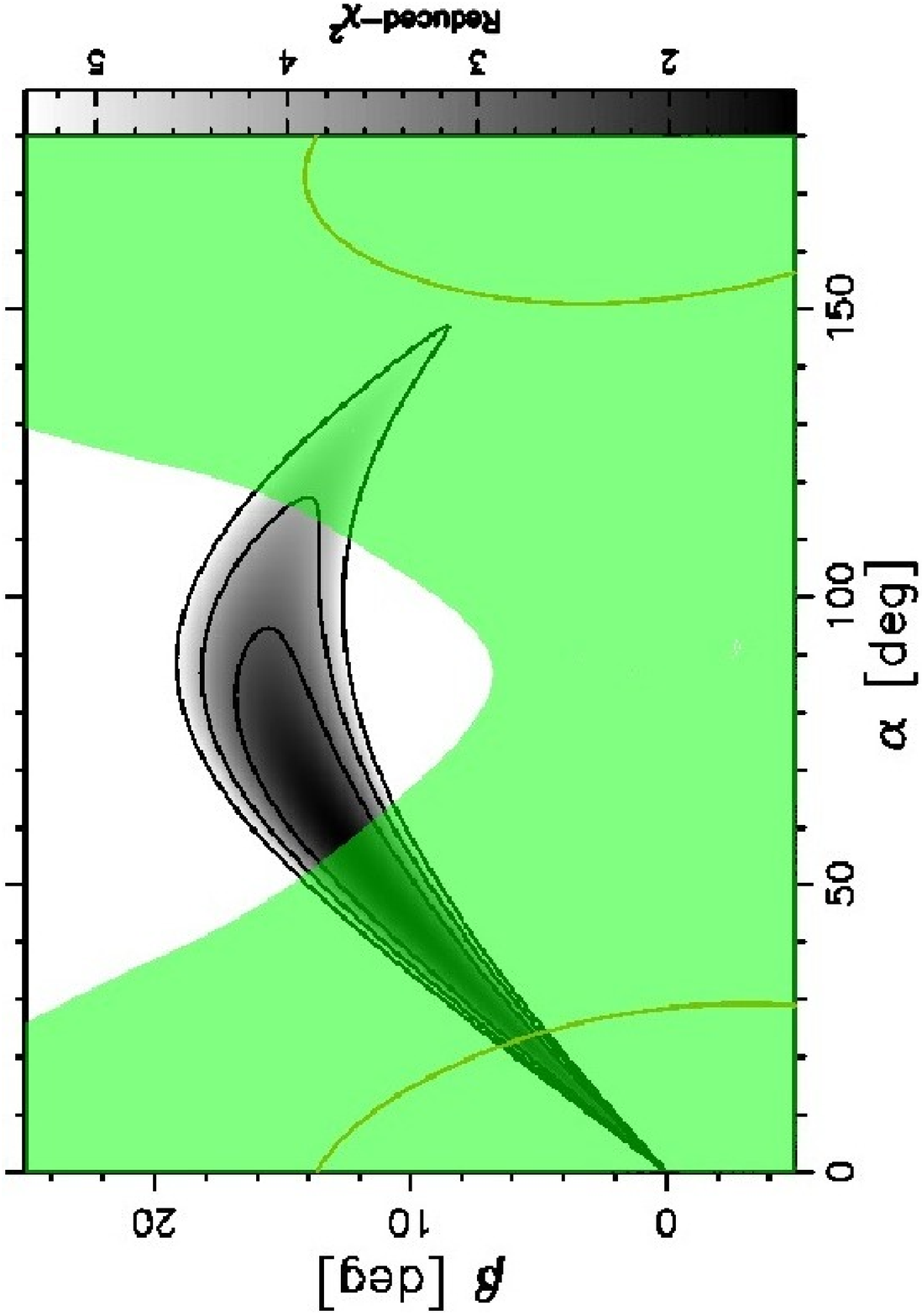} 
\caption{\label{FigJ1718-3825} PSR J1718--3825 at 20 cm. As Fig.~\ref{FigJ0631+1036}. } 
\end{figure}

\subsection{PSR J1718--3825 (Fig.~\ref{FigJ1718-3825})} 
 
The profile exhibits three components. The central component could be a core component, with a conal component on either side. Alternatively, the profile could show one side of a core-double-cone configuration, in which case either of the outer components could be the core. To allow for these possibilities, positions of the fiducial plane between the leading and trailing components were considered. The midpoint between the peaks of the two outer components, which roughly coincides with the centre of the profile, was taken as the favoured position of the fiducial plane. The large width of the profile suggests that the axes are relatively aligned and that $\alpha < 63\degree$ or $111\degree < \alpha < 148\degree$.

This pulsar was included in the sample of \cite{waa+10} and the results
obtained here are largely consistent with theirs.
 
\begin{figure} 
\centering 
\includegraphics[height=0.93\hsize,angle=270]{J1730-3350_paswing.ps} 
\includegraphics[height=\hsize,angle=270]{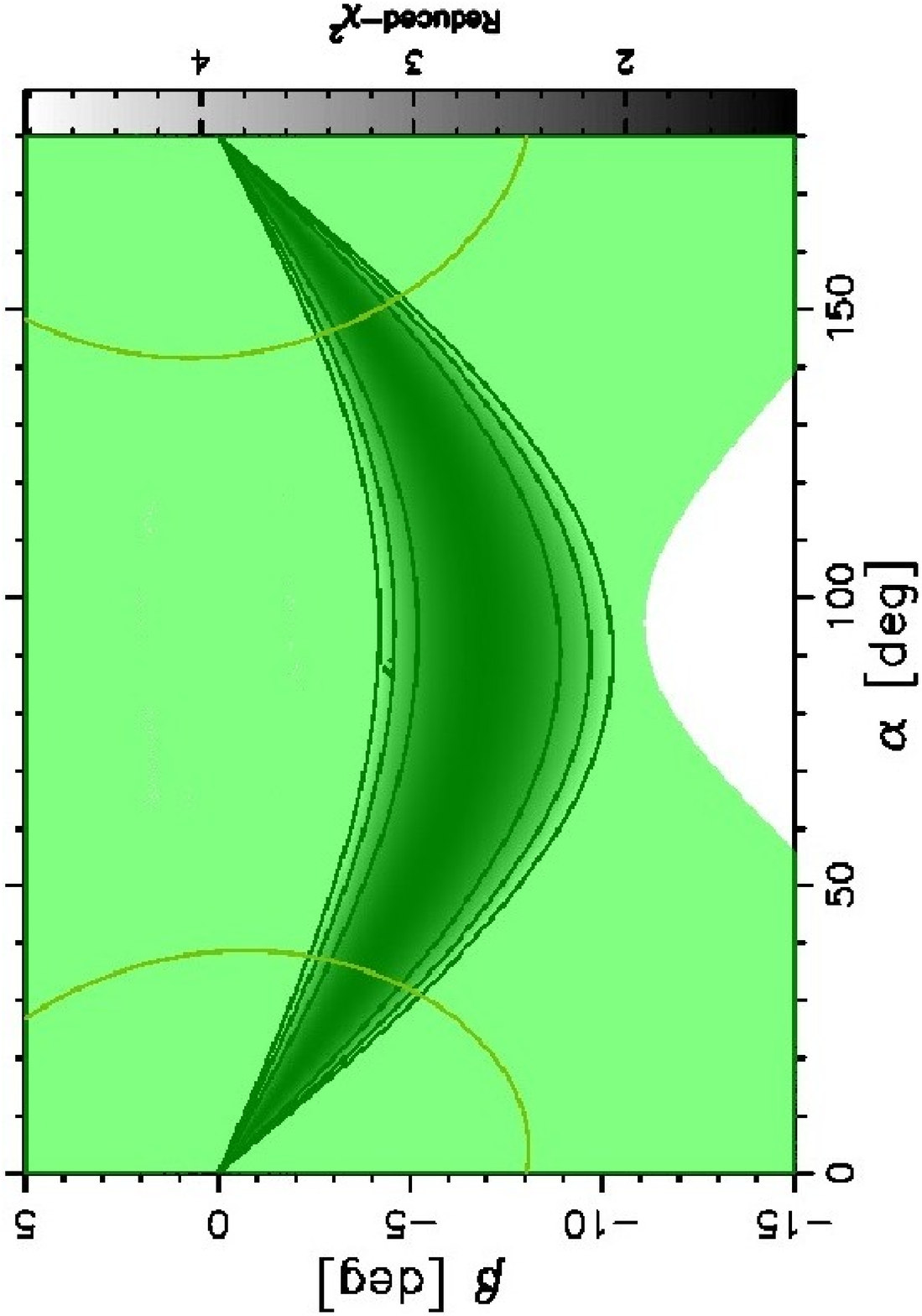} 
\caption{\label{FigJ1730-3350} PSR J1730--3350 at 10 cm. As Fig.~\ref{FigJ0631+1036}. } 
\end{figure}

\subsection{PSR J1730--3350 / B1727--33 (Fig.~\ref{FigJ1730-3350})} 
 
The data used here were taken at 10 cm as the profile is scattered at longer wavelengths. The figure shows that the RVM fit for this pulsar can constrain $-10\degree < \beta < 0$, but leaves $\alpha$ unconstrained. If the fiducial plane coincided with the single peak in the profile it would clearly be later than the inflection point, which would result in an unphysical negative emission height. 

Furthermore, the observed component is steeper at the trailing edge, which is characteristic of a leading conal component. This could suggest that any missing component would be later, not earlier, than the observed component. However, to prevent a negative emission height it seems likely that there is an absent leading component (similar to the transient trailing component of PSR J1119--6127, see $\S$~\ref{J1119}). This would allow the fiducial plane to be at a sufficiently early phase to make the offset $\Delta\phi$ positive and is therefore included in the allowed range for $\phi_{\mathrm{fid}}$. The inferred values of $\alpha$ and $\beta$ are highly sensitive to the fiducial plane position and the conservative range chosen here does not allow the viewing geometry to be constrained further than the $\chi^2$ surface.

\cite{cmk01} reported a constraint of $|\beta| < 5\degree$ for this pulsar from RVM fitting. However, the data they used were taken at 1351 MHz. Hence, scattering will have affected their RVM fit and is likely to be the source of the discrepancy between that result and the constraint presented here.

\begin{figure} 
\centering 
\includegraphics[height=0.93\hsize,angle=270]{J1801-2451_paswing.ps} 
\includegraphics[height=\hsize,angle=270]{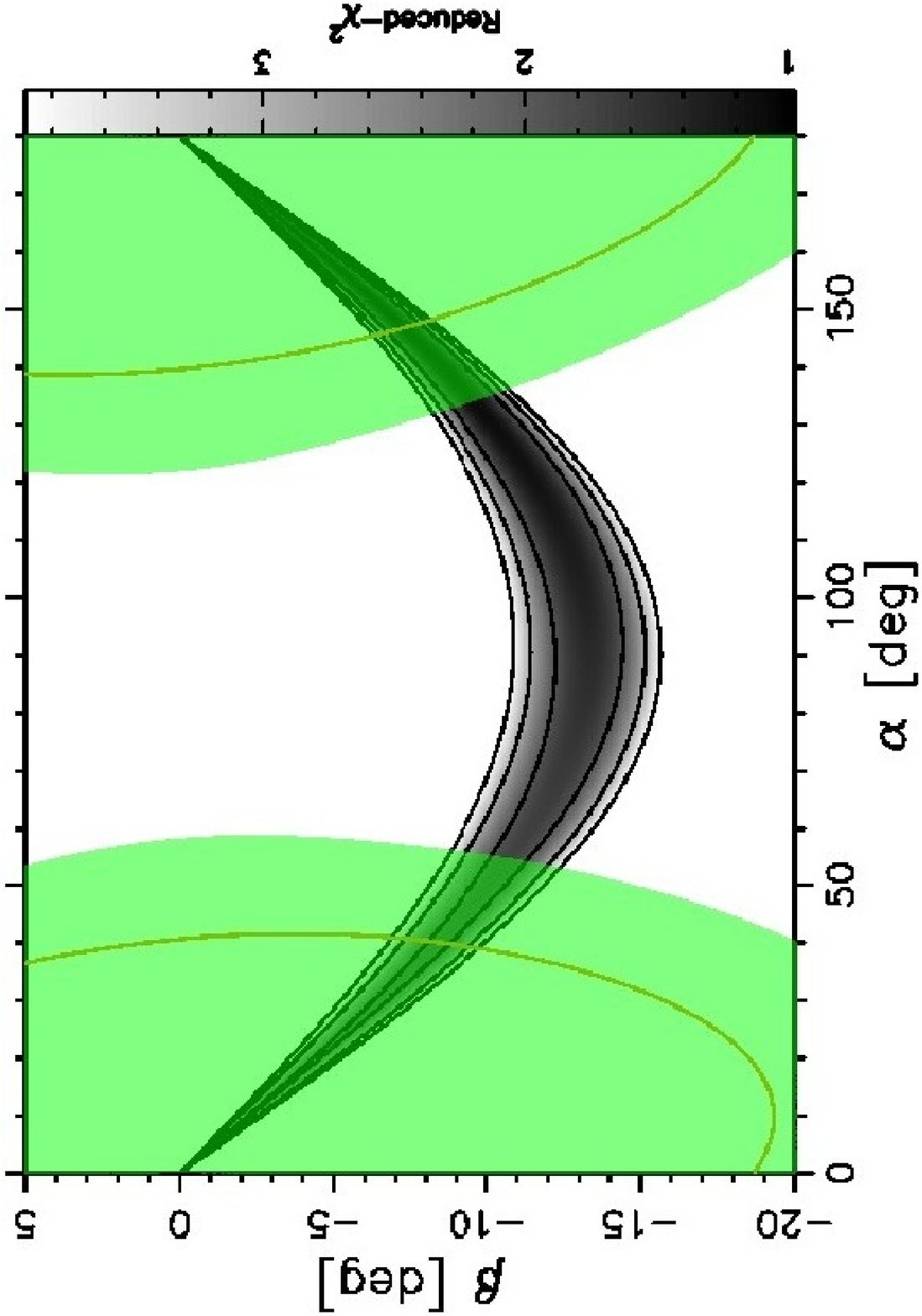} 
\caption{\label{FigJ1801-2451} PSR J1801--2451 at 20 cm. As Fig.~\ref{FigJ0631+1036}. } 
\end{figure}

\subsection{PSR J1801--2451 / B1757--24 (Fig.~\ref{FigJ1801-2451})} 
 
The profile shows a single component, suggesting the fiducial plane could be close to the peak. Alternatively, this may be a trailing conal component; the profile is slightly steeper at the leading edge, which is characteristic of such components. In this case the fiducial plane would be close to the leading edge of the pulse. The inflection point of the best RVM fit is slightly earlier in phase than the peak, although the relatively large 3$\sigma$ error on this value allows small positive offsets even if $\phi_{\mathrm{fid}}$ is close to the peak. The combined constraint from RVM fitting and the emission height limits $\alpha < 57\degree$ or $> 131\degree$ and $-12\degree < \beta < 0$.

\begin{figure} 
\centering 
\includegraphics[height=0.93\hsize,angle=270]{J1835-1106_paswing.ps} 
\includegraphics[height=\hsize,angle=270]{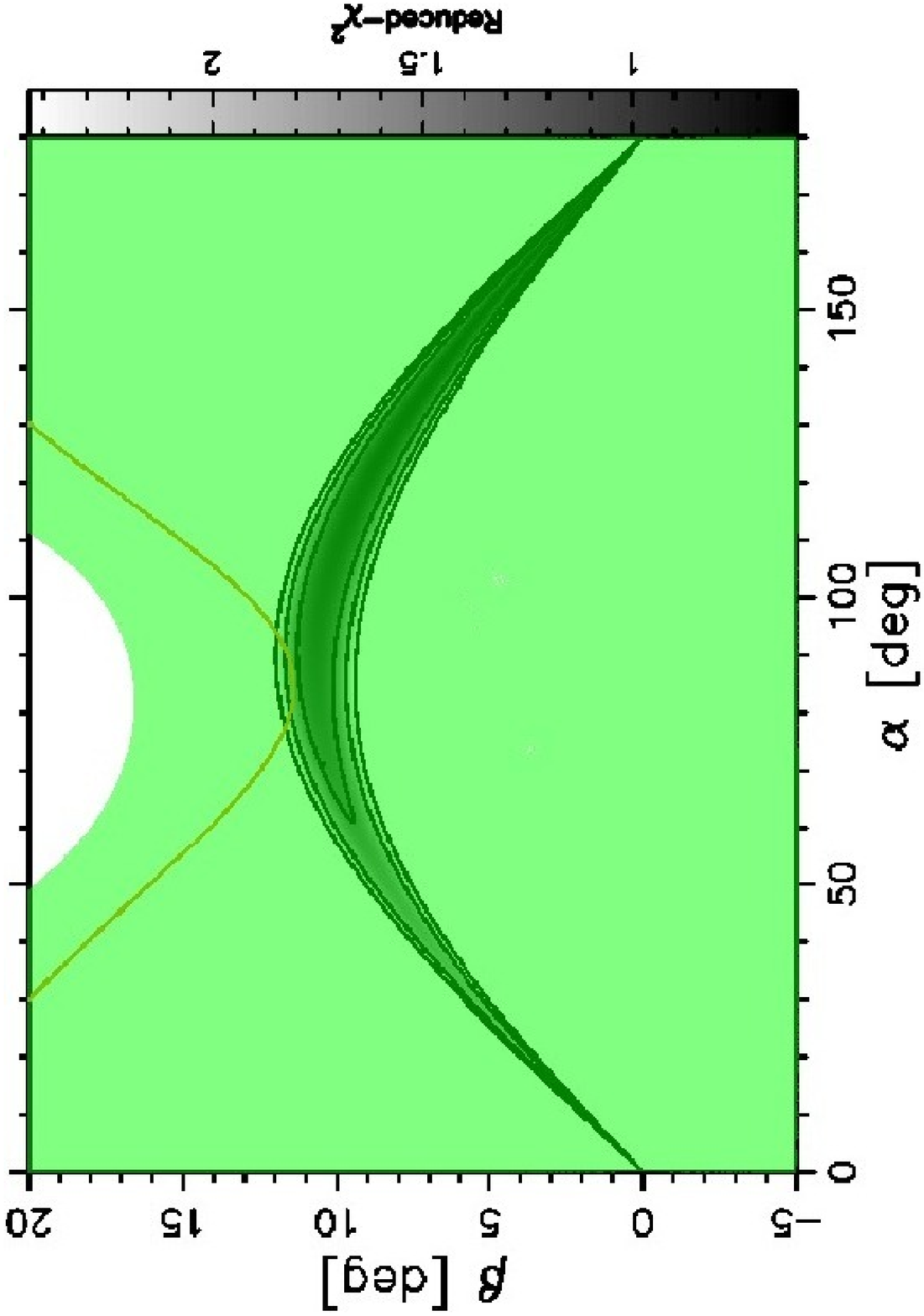} 
\caption{\label{FigJ1835-1106} PSR J1835--1106 at 20 cm. As Fig.~\ref{FigJ0631+1036}. } 
\end{figure}

\subsection{PSR J1835--1106 (Fig.~\ref{FigJ1835-1106})} 
 
The $\chi^2$ surface indicates that $0\degree < \beta < 13\degree$. The profile exhibits a single peak, which could indicate that the fiducial plane is at the observed peak. However, it was noted by JW06 that the trailing component of a double profile in young pulsars is often the more intense. It is therefore possible that there is a leading component which has not been observed. The range of fiducial plane positions was chosen to reflect these two possibilities. This large $\phi_{\mathrm{fid}}$ range, along with the comparatively large error on $\phi_0$, leads to a large range of possible $\rho$ values. As a result the A/R effect does not help constrain $\alpha$.

\begin{table*}
\footnotesize
\centering
\setlength{\extrarowheight}{1 mm}
\caption{\label{TabDerivedParameters} Values of the key measured and derived parameters of each pulsar in the sample, as determined during the process of constraining the viewing geometry. These are, respectively, the observation wavelength ($\lambda$), the rotation measure used to de-Faraday rotate the data, the pulse phase resolution ($N_{\mathrm{bins}}$) used, the phase of the inflection point, the phase of the fiducial plane, the relative offset, the inferred range of rotational phase for which the line of sight samples the open-field-line region, $W_{\mathrm{open}}$, and the derived emission height. In the cases of PSRs J0908--4913 and J1057--5226, MP and IP refer to the main pulse and interpulse. In the case of PSR J1119--6127, (a) and (b) correspond to the scenarios in which the RRAT-like components are outside and inside the open-field-line region respectively (see $\S$~\ref{J1119}). Rotation measure references: [1] \protect \cite{njk+08}; [2] \protect \cite{wck+04}; [3] \protect \cite{jw06}; [4] \protect \cite{jhv+05}; [5] \protect \cite{qlm+95}; [6] \protect \cite{hml+06}; [7] \protect \cite{kjk+08}; [8] \protect \cite{tml93}; [9] \protect \cite{ojk+08}; [10] \protect \cite{wj08b}; [11] \protect \cite{cmk01}. }

\begin{tabular}{lcrcr@{ $\pm$ }lr@{ $\pm$ }lr@{ $\pm$ }lr@{ $\pm$ }lr@{ $\pm$ }l}
\hline
PSR & $\lambda$ & \multicolumn{1}{c}{RM} & $N_{\mathrm{bins}}$ & \multicolumn{2}{c}{$\phi_0$ / $\degree$} & \multicolumn{2}{c}{$\phi_{\mathrm{fid}}$ / $\degree$} & \multicolumn{2}{c}{$\Delta\phi$ / $\degree$} & \multicolumn{2}{c}{$W_{\mathrm{open}}$ / $\degree$} & \multicolumn{2}{c}{$h_{\mathrm{em}}$ / km} \\
& \multicolumn{1}{c}{/ cm} & / rad $\mathrm{m}^{-2}$ & / bins & \multicolumn{2}{c}{} & \multicolumn{2}{c}{} & \multicolumn{2}{c}{} & \multicolumn{2}{c}{} & \multicolumn{2}{c}{} \\
\hline

J0631+1036 & 20 & 137\hspace{10 pt} $^{[1]}$ & \hspace{3 pt}512 & 185.4 & $ ^{3.6}_{1.6}$ & 177.0 & 3.0 & 8.4 & $ ^{6.6}_{4.6}$ & 26 & $ ^{12}_{5}$ & 500 & $ ^{400}_{280}$ \\
J0659+1414 & 20 & 23.5\hspace{4 pt} $^{[2]}$ & \hspace{3 pt}128 & 193.7 & $ ^{8.6}_{4.6}$ & 178.0 & $ ^{5.0}_{23.0}$ & 15.7 & $ ^{31.6}_{9.6}$ & 34 & $ ^{60}_{7}$ & 1300 & $ ^{2500}_{800}$ \\
J0729--1448 & 20 & 46\hspace{10 pt} $^{[3]}$ & \hspace{3 pt}256 & 184.4 & $ ^{1.0}_{1.1}$ & 173.0 & $ ^{7.0}_{8.0}$ & 11.4 & $ ^{9.0}_{8.1}$ & 29 & $ ^{19}_{7}$ & 600 & $ ^{470}_{420}$ \\
J0742--2822 & 20 & 149.95 $^{[4]}$ & 1024 & 188.6 & $ ^{1.5}_{0.4}$ & 183.0 & $ ^{7.0}_{4.0}$ & 5.6 & $ ^{5.5}_{5.6}$ & 16 & $ ^{14}_{2}$ & 190 & $190$ \\
J0835--4510 & 20 & 31.38 $^{[4]}$ & 1024 & 184.4 & $ ^{0.4}_{0.3}$ & 180.0 & 2.0 & 4.4 & $ ^{2.4}_{2.3}$ & 20 & 9 & 81 & $ ^{44}_{42}$ \\
J0908--4913 (MP) & 20 & 10\hspace{10 pt} $^{[5]}$ & 1024 & 94.3 & $ ^{0.4}_{0.5}$ & 81.0 & $ ^{9.0}_{0.0}$ & 13.3 & $ ^{0.4}_{9.5}$ & 24 & $ ^{1}_{14}$ & 300 & $ ^{10}_{210}$ \\
J0908--4913 (IP) & 20 & 10\hspace{10 pt} $^{[5]}$ & 1024 & 274.3 & $ ^{0.4}_{0.5}$ & 263.0 & $ ^{5.5}_{6.0}$ & 11.3 & $ ^{6.4}_{6.0}$ & 18 & $ ^{13}_{3}$ & 250 & $ ^{140}_{130}$ \\
J0940--5428 & 20 & --31\hspace{10 pt} $^{[3]}$ & \hspace{3 pt}256 & 179.7 & $ ^{4.3}_{10.9}$ & 171.0 & $ ^{10.0}_{8.0}$ & 8.7 & $ ^{12.3}_{8.7}$ & 48 & $ ^{20}_{14}$ & 160 & $ ^{230}_{160}$ \\
J1016--5857 & 20 & --540\hspace{10 pt} $^{[3]}$ & \hspace{3 pt}256 & 165.1 & $ ^{4.6}_{5.7}$ & 162.0 & $ ^{6.0}_{4.0}$ & 3.1 & $ ^{8.6}_{3.1}$ & 61 & $ ^{11}_{15}$ & 69 & $ ^{191}_{69}$ \\
J1019--5749 & 10 & --366\hspace{10 pt} $^{[6]}$ & \hspace{3 pt}128 & 163.2 & 5.8 & \multicolumn{2}{c}{---} & \multicolumn{2}{c}{---} & \multicolumn{2}{c}{---} & \multicolumn{2}{c}{---} \\
J1028--5820 & 20 & --5\hspace{10 pt} $^{[7]}$ & 1024 & \multicolumn{2}{c}{---} & 179.5 & 1.0 & \multicolumn{2}{c}{---} & 3 & $ ^{2}_{0}$ & \multicolumn{2}{c}{---} \\
J1048--5832 & 20 & --155\hspace{10 pt} $^{[5]}$ & \hspace{3 pt}256 & 183.2 & $ ^{0.9}_{0.7}$ & 180.0 & $ ^{6.0}_{19.0}$ & 3.2 & $ ^{19.9}_{3.2}$ & 34 & $ ^{40}_{11}$ & 82 & $ ^{513}_{82}$ \\
J1057--5226 (MP) & 20 & 47.2\hspace{4 pt} $^{[8]}$ & 1024 & 90.6 & $ ^{8.8}_{14.9}$ & 90.0 & $ ^{22.0}_{14.0}$ & 0.6 & $ ^{22.8}_{0.6}$ & 46 & $ ^{29}_{14}$ & 24 & $ ^{935}_{24}$ \\
J1057--5226 (IP) & 20 & 47.2\hspace{4 pt} $^{[8]}$ & 1024 & 270.6 & $ ^{8.8}_{14.9}$ & 255.6 & $ ^{8.4}_{14.6}$ & 15.0 & $ ^{23.4}_{15.0}$ & 45 & $ ^{32}_{9}$ & 620 & $ ^{960}_{620}$ \\
J1105--6107 & 20 & 187\hspace{10 pt} $^{[3]}$ & \hspace{3 pt}256 & 183.3 & $ ^{0.9}_{0.6}$ & 174.0 & $ ^{6.0}_{7.0}$ & 9.3 & $ ^{7.9}_{6.6}$ & 25 & $ ^{17}_{4}$ & 120 & $ ^{100}_{90}$ \\
J1112--6103 & 10 & 242\hspace{10 pt} $^{[6]}$ & \hspace{3 pt}256 & 182.8 & $ ^{1.4}_{1.0}$ & 178.0 & $ ^{6.0}_{9.0}$ & 4.8 & $ ^{10.4}_{4.8}$ & 29 & $ ^{20}_{4}$ & 64 & $ ^{140}_{64}$ \\
J1119--6127 (a) & 20 & 853\hspace{10 pt} $^{[3]}$ & \hspace{3 pt}128 & 190.3 & $ ^{18.8}_{6.2}$ & 188.5 & $ ^{10.0}_{11.5}$ & 1.8 & $ ^{30.3}_{1.8}$ & 72 & $ ^{25}_{12}$ & 150 & $ ^{2570}_{150}$ \\
J1119--6127 (b) & 20 & 853\hspace{10 pt} $^{[3]}$ & \hspace{3 pt}128 & 190.3 & $ ^{18.8}_{6.2}$ & 188.5 & $ ^{10.0}_{11.5}$ & 1.8 & $ ^{30.3}_{1.8}$ & 102 & $ ^{25}_{17}$ & 150 & $ ^{2570}_{150}$ \\
J1357--6429 & 20 & --47\hspace{10 pt} $^{[3]}$ & \hspace{3 pt}128 & 181.7 & $ ^{39.4}_{4.6}$ & 180.0 & $ ^{10.0}_{35.0}$ & 1.7 & $ ^{74.4}_{1.7}$ & 72 & $ ^{82}_{20}$ & 58 & $ ^{2571}_{58}$ \\
J1410--6132 & \hspace{1 pt} 5 & 2400\hspace{10 pt} $^{[9]}$ & \hspace{3 pt}128 & 177.1 & $ ^{0.6}_{0.7}$ & 176.0 & $ ^{5.0}_{6.0}$ & 1.1 & $ ^{6.6}_{1.1}$ & 20 & $ ^{16}_{8}$ & 11 & $ ^{68}_{11}$ \\
J1420--6048 & 20 & --122\hspace{10 pt} $^{[3]}$ & \hspace{3 pt}128 & 167.7 & $ ^{2.1}_{2.5}$ & 163.0 & $ ^{20.0}_{21.0}$ & 4.7 & $ ^{23.1}_{4.7}$ & 74 & $ ^{52}_{16}$ & 66 & $ ^{327}_{66}$ \\
J1509--5850 & 20 & 0\hspace{21 pt} & \hspace{3 pt}256 & \multicolumn{2}{c}{---} & 191.0 & $ ^{15.0}_{14.0}$ & \multicolumn{2}{c}{---} & 52 & $ ^{36}_{14}$ & \multicolumn{2}{c}{---} \\
J1513--5908 & 20 & 216\hspace{10 pt} $^{[3]}$ & \hspace{3 pt}128 & 172.4 & $ ^{105.6}_{104.4}$ & 155.0 & $ ^{32.0}_{35.0}$ & 17.4 & $ ^{140.6}_{17.4}$ & 114 & $ ^{82}_{60}$ & 550 & $ ^{4420}_{550}$ \\
J1531--5610 & 20 & --50\hspace{10 pt} $^{[6]}$ & \hspace{3 pt}256 & 178.7 & $ ^{33.0}_{8.8}$ & 170.0 & $ ^{11.0}_{10.0}$ & 8.7 & $ ^{43.0}_{8.7}$ & 38 & $ ^{36}_{11}$ & 150 & $ ^{750}_{150}$ \\
J1648--4611 & 10 & --682\hspace{10 pt} $^{[6]}$ & \hspace{3 pt}512 & 185.1 & $ ^{7.1}_{9.8}$ & 184.0 & $ ^{5.0}_{6.0}$ & 1.1 & $ ^{13.1}_{1.1}$ & 22 & $ ^{14}_{4}$ & 37 & $ ^{450}_{37}$ \\
J1702--4128 & 10 & --160\hspace{6 pt} $^{[10]}$ & \hspace{3 pt}256 & 177.9 & $ ^{4.9}_{4.5}$ & 176.0 & $ ^{5.0}_{18.0}$ & 1.9 & $ ^{22.9}_{1.9}$ & 34 & $ ^{42}_{11}$ & 71 & $ ^{867}_{71}$ \\
J1709--4429 & 20 & 0.7\hspace{4 pt} $^{[4]}$ & \hspace{3 pt}128 & 190.9 & $ ^{3.2}_{2.5}$ & 180.0 & $ ^{5.0}_{35.0}$ & 10.9 & $ ^{38.2}_{7.5}$ & 46 & $ ^{74}_{11}$ & 230 & $ ^{810}_{160}$ \\
J1718--3825 & 20 & 113\hspace{10 pt} $^{[6]}$ & \hspace{3 pt}256 & 203.7 & $ ^{7.4}_{4.4}$ & 198.0 & $ ^{17.0}_{18.0}$ & 5.7 & $ ^{25.4}_{5.7}$ & 58 & $ ^{52}_{12}$ & 89 & $ ^{396}_{89}$ \\
J1730--3350 & 10 & --142\hspace{6 pt} $^{[11]}$ & \hspace{3 pt}512 & 175.0 & $ ^{3.5}_{2.6}$ & 173.0 & $ ^{6.0}_{5.0}$ & 2.0 & $ ^{8.5}_{2.0}$ & 26 & $ ^{16}_{18}$ & 57 & $ ^{245}_{57}$ \\
J1801--2451 & 20 & 637\hspace{10 pt} $^{[6]}$ & \hspace{3 pt}256 & 176.6 & $ ^{2.8}_{5.7}$ & 166.0 & $ ^{16.0}_{4.0}$ & 10.6 & $ ^{6.8}_{10.6}$ & 58 & $ ^{16}_{38}$ & 280 & $ ^{180}_{280}$ \\
J1835--1106 & 20 & 42\hspace{10 pt} $^{[1]}$ & \hspace{3 pt}512 & 186.0 & $ ^{1.7}_{6.6}$ & 169.0 & $ ^{13.0}_{5.0}$ & 17.0 & $ ^{6.7}_{17.0}$ & 42 & $ ^{14}_{27}$ & 590 & $ ^{230}_{590}$ \\

\hline

\end{tabular}
\end{table*}

\begin{table*}
\footnotesize
\centering
\setlength{\extrarowheight}{1 mm}
\caption{\label{TabUncorrectedViewingGeometries} The beam half-opening angles, allowed and favoured viewing geometries for the sample. In the cases of PSRs J0908--4913 and J1057--5226, MP and IP refer to $\rho$ values for, and $\alpha$ and $\beta$ values with respect to, the main pulse and interpulse. In the case of PSR J1119--6127, (a) and (b) correspond to the scenarios in which the RRAT-like components are outside and inside the open-field-line region respectively. Footnotes: * The favoured $\alpha$ value is set to 90$\degree$ as these pulsars are argued to be orthogonal rotators (see text). The favoured $\beta$ is that at which the reduced-$\chi^2$ is lowest for $\alpha = 90\degree$. $^{\dagger}$ The allowed ranges of $\alpha$ and $\beta$ were derived from our data whereas the given favoured solution is that determined at 8.4~GHz by \protect \cite{kj08}, as discussed in $\S$~\ref{J0908}. $^{\triangle}$ The beamwidth and viewing geometry could not be calculated as the relative offset of the inflection point and fiducial plane could not be reliably determined. $^{\ddagger}$ The $\alpha$ and $\beta$ values were derived using the A/R effect for the interpulse only (see $\S$~\ref{J1057}). }

\begin{tabular}{lr@{ $\pm$ }l>{\hspace{6 mm}}r@{ - }lr@{ - }lrr}
\hline
PSR & \multicolumn{2}{c}{} & \multicolumn{4}{c}{Allowed Solutions} & \multicolumn{2}{c}{Favoured Solutions} \\
& \multicolumn{2}{c}{$\rho$ / $\degree$} & \multicolumn{2}{c}{$\alpha$ / $\degree$} & \multicolumn{2}{c}{$\beta$ / $\degree$} & \multicolumn{1}{c}{$\alpha$ / $\degree$} & $\beta$ / $\degree$ \\
\hline
J0631+1036 & 16.61 & $ ^{5.75}_{5.49}$ & 33 & 152 & --10.5 & --2.5 & 97.2\hspace{7 pt} & --10.4\hspace{7 pt} \\
J0659+1414 & 22.89 & $ ^{18.43}_{8.78}$ & 32 & 160 & --22 & --5 & 120.7\hspace{7 pt} & --16.7\hspace{7 pt} \\
J0729--1448 & 19.41 & $ ^{6.83}_{9.06}$ & 32 & 148 & 2 & 7 & 90\hspace{7 pt} * & 6\hspace{6 pt} * \\
J0742--2822 & 13.52 & $ ^{5.63}_{13.52}$ & 55 & 180 & --7 & 0 & 90\hspace{7 pt} * & --6.5 * \\
J0835--4510 & 11.97 & $ ^{2.95}_{3.72}$ & 40 & 98 & --7.5 & --5 & 74.2\hspace{7 pt} & --7.3\hspace{7 pt} \\
J0908--4913 (MP) & 21.01 & $ ^{0.32}_{9.9}$ & 96 & 96.8 & --8.5 & --6.3 & 96.1 $^{\dagger}$ & --5.9 $^{\dagger}$ \\
J0908--4913 (IP) & 19.33 & $ ^{5.04}_{6.18}$ & \multicolumn{2}{c}{} & \multicolumn{2}{c}{} & 83.9 $^{\dagger}$ & 6.3 $^{\dagger}$ \\
J0940--5428 & 16.91 & $ ^{9.74}_{16.91}$ & \multicolumn{2}{c}{0 - 49; 122 - 180} & 0 & 18 & 143.5\hspace{7 pt} & 12.0\hspace{7 pt} \\
J1016--5857 & 10.03 & $ ^{9.64}_{10.03}$ & \multicolumn{2}{c}{0 - 39; 145 - 180} & --8.5 & 0 & 163.0\hspace{7 pt} & --3.1\hspace{7 pt} \\
J1019--5749 $^{\triangle}$ & \multicolumn{2}{c}{---} & \multicolumn{2}{c}{---} & \multicolumn{2}{c}{---} & ---\hspace{10 pt} & ---\hspace{10 pt} \\
J1028--5820 $^{\triangle}$ & \multicolumn{2}{c}{---} & \multicolumn{2}{c}{---} & \multicolumn{2}{c}{---} & ---\hspace{10 pt} & ---\hspace{10 pt} \\
J1048--5832 & 10.19 & $ ^{17.82}_{10.19}$ & 0 & 50 & 0 & 7.5 & 29.6\hspace{7 pt} & 4.8\hspace{7 pt} \\
J1057--5226 (MP) & 4.4 & $ ^{23.8}_{4.4}$ & 68 & 98 & 8 & 48 & 75.6 $^{\ddagger}$ & 34.7 $^{\ddagger}$ \\
J1057--5226 (IP) & 22.36 & $ ^{14.45}_{22.36}$ & \multicolumn{2}{c}{} & \multicolumn{2}{c}{} & 104.4 $^{\ddagger}$ & 5.9 $^{\ddagger}$ \\
J1105--6107 & 17.49 & $ ^{6.51}_{8.14}$ & 28 & 140 & 2 & 5 & 90\hspace{7 pt} * & 4\hspace{6 pt} * \\
J1112--6103 & 12.5 & $ ^{10.01}_{12.5}$ & 0 & 180 & --5.5 & 0 & 126.6\hspace{7 pt} & --3.8\hspace{7 pt} \\
J1119--6127 (a) & 7.63 & $ ^{25.75}_{7.63}$ & \multicolumn{2}{c}{0 - 62; 132 - 180} & --22 & 0 & 170.9\hspace{7 pt} & --3.8\hspace{7 pt} \\
J1119--6127 (b) & 7.63 & $ ^{25.75}_{7.63}$ & \multicolumn{2}{c}{0 - 48; 144 - 180}    & --19 & 0 & 173.1\hspace{7 pt} & --2.9\hspace{7 pt} \\
J1357--6429 & 7.42 & $ ^{47.19}_{7.42}$ & \multicolumn{2}{c}{0 - 55; 102 - 180} & 0 & 50 & 6.9\hspace{7 pt} & 4.9\hspace{7 pt} \\
J1410--6132 & 5.96 & $ ^{9.93}_{5.96}$ & 0 & 180 & 0 & 5.5 & 147.0\hspace{7 pt} & 2.9\hspace{7 pt} \\
J1420--6048 & 12.37 & $ ^{18.53}_{12.37}$ & 0 & 33 & 0 & 8.5 & 17.1\hspace{7 pt} & 3.8\hspace{7 pt} \\
J1509--5850 $^{\triangle}$ & \multicolumn{2}{c}{---} & \multicolumn{2}{c}{---} & \multicolumn{2}{c}{---} & ---\hspace{10 pt} & ---\hspace{10 pt} \\
J1513--5908 & 24.15 & $ ^{68.66}_{24.15}$ & 0 & 180 & 0 & 70 & 13.7\hspace{7 pt} & 15.3\hspace{7 pt} \\
J1531--5610 & 16.91 & $ ^{26.55}_{16.91}$ & 0 & 180 & --43 & 0 & 158.4\hspace{7 pt} & --14.4\hspace{7 pt} \\
J1648--4611 & 5.96 & $ ^{15.77}_{5.96}$ & 0 & 180 & --14 & 0 & 158.9\hspace{7 pt} & --4.1\hspace{7 pt} \\
J1702--4128 & 7.84 & $ ^{21.24}_{7.84}$ & \multicolumn{2}{c}{0 - 68; 120 - 180} & --13.5 & 0 & 22.8\hspace{7 pt} & --5.3\hspace{7 pt} \\
J1709--4429 & 18.97 & $ ^{23.23}_{8.46}$ & 12 & 50 & 5.5 & 19 & 31.6\hspace{7 pt} & 12.9\hspace{7 pt} \\
J1718--3825 & 13.64 & $ ^{19.18}_{13.64}$ & \multicolumn{2}{c}{0 - 63; 111 - 148} & 0 & 16 & 23.1\hspace{7 pt} & 5.5\hspace{7 pt} \\
J1730--3350 & 8.05 & $ ^{10.57}_{8.05}$ & 0 & 180 & --10 & 0 & 148.4\hspace{7 pt} & --3.7\hspace{7 pt} \\
J1801--2451 & 18.7 & $ ^{5.45}_{18.7}$ & \multicolumn{2}{c}{0 - 57; 131 - 180} & --12 & 0 & 146.8\hspace{7 pt} & --7.3\hspace{7 pt} \\
J1835--1106 & 23.86 & $ ^{4.54}_{23.86}$ & 0 & 180 & 0 & 13 & 85.5\hspace{7 pt} & 11.4\hspace{7 pt} \\
\hline

\end{tabular}
\end{table*}

\section{The derived $\alpha$ distribution}
\label{SectAlphaDistribution}

The $\alpha$ distribution of the sample was investigated by considering the favoured $\alpha$ values of 25 of the pulsars. These and the associated $\beta$ values are given in Table~\ref{TabUncorrectedViewingGeometries}. Each favoured value was determined by finding the ``crossing point'', the point along the favoured contour ($\S$~\ref{SectEmissionHeightConstraint}) for which $\chi^2$ was lowest and within the 3$\sigma$ limit. In three out of 28 cases this was not possible. These were PSR J1019--5749, for which $\phi_{\mathrm{fid}}$ could not be determined due to scattering-induced distortion of the profile, 
and PSRs J1028--5819 and J1509--5850, for which $\phi_0$ could not be determined due to the shallow gradient of the respective PA curves.  In four cases\footnote{PSRs~J0729--1448, J0742--2822, J0908--4913 and J1105--6107} the favoured contour was inconsistent with the $\chi^2$ surface. Possible reasons for this are a small error in the choice of the preferred position of the fiducial plane or measurement uncertainties in the position of the inflection point or pulse edges. Alternatively it is possible that both intersections of the last-open-field-lines by the line of sight occur outside the observed pulse, in which case $W_{\mathrm{open}}$ will have been underestimated. It follows from Eq.~\ref{EqRhoContours} that a given $\rho$ contour will deform and shift towards larger $|\beta|$ values. For these pulsars the favoured contour was closest to the 3$\sigma$ limit when $\alpha \approx 90\degree$, so these pulsars were assumed to be orthogonal rotators. In the table we use the result from \cite{kj08} for PSR~J0908--4913 and set $\alpha = 90\degree$ for the other three pulsars.

\begin{figure}
\centering
\includegraphics[height=\hsize,angle=270]{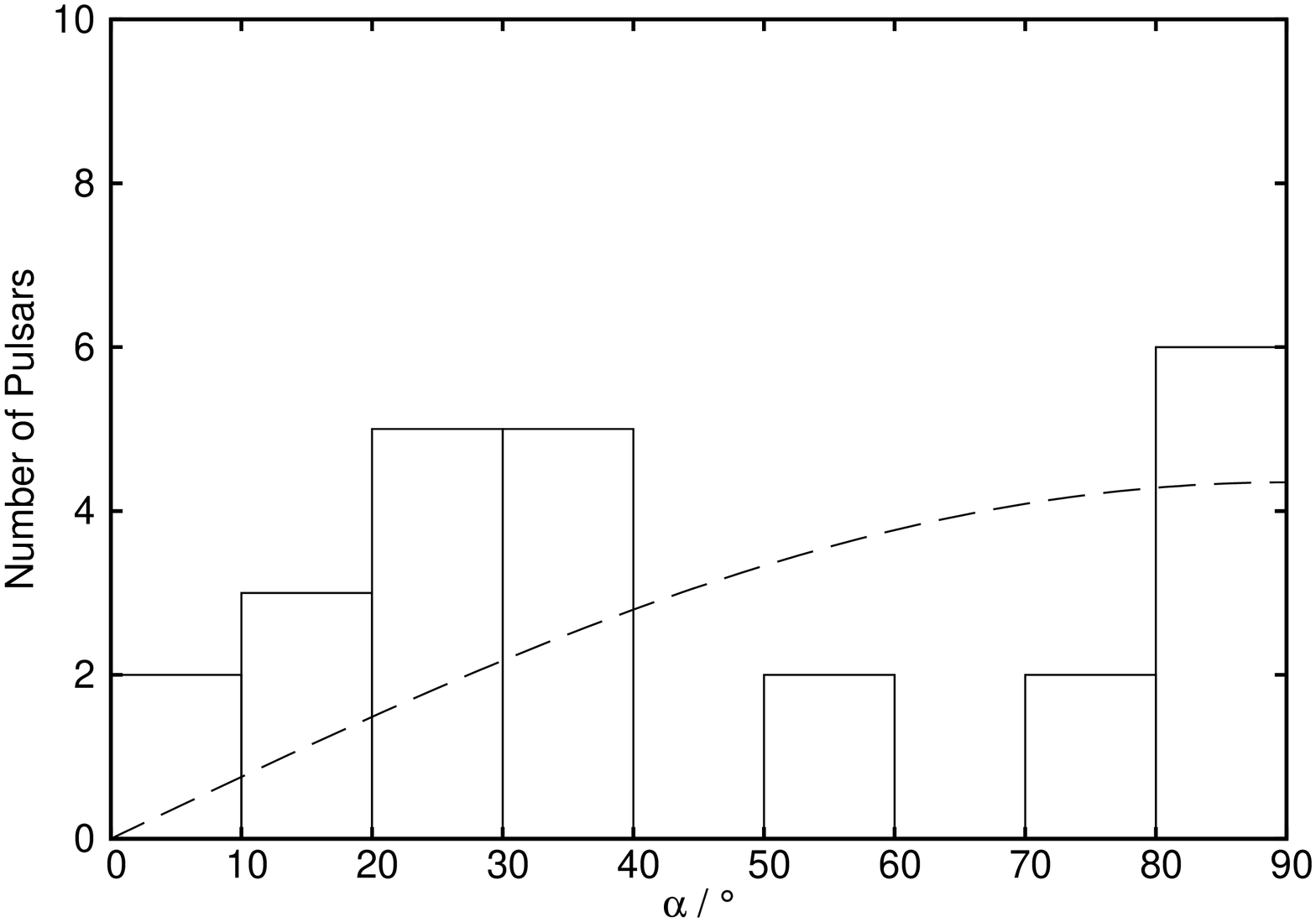}
\caption{\label{FigOriginalDistribution} Distribution of most likely $\alpha$ values (mapped into the interval $0\degree < \alpha < 90\degree$). The sinusoidal distribution which might be expected for a sample of young pulsars is shown as the dashed curve. A preference for low $\alpha$ values relative to the sinusoidal distribution can be clearly seen.  }
\end{figure}

The resulting $\alpha$ distribution (Fig.~\ref{FigOriginalDistribution}) shows a pronounced skew towards low $\alpha$ values, with an unexpectedly low number of sources with $40\degree < \alpha < 80\degree$. In a typical sample of young pulsars, for which alignment of the axes via magnetic torques has yet to take effect \citep{tm98, wj08a, ycb+10}, the axes would be expected to be randomly orientated, leading to a sinusoidal $\alpha$ distribution (e.g., \citealt{gh96}). Furthermore, our sample of young pulsars are all $\gamma$-ray-detected. \cite{wrw+09} predicted that pulsars are more easily detectable in $\gamma$-rays when $\alpha$ is large, and therefore we would expect our sample to have a bias towards high $\alpha$ values relative to the sinusoidal distribution. The figure shows that this clearly is not the case. The Kolmogorov-Smirnov (KS) test allows us to quantify how dissimilar two distributions are \citep{pftv86}. A low enough probability indicates that the two distributions are statistically different. The result of such a test in which the observed $\alpha$ distribution was compared to a sinusoidal distribution was 0.037\%, indicating to a confidence greater than 3 $\sigma$ that the observed values are not drawn from a sinusoidal distribution\footnote{To check the validity of this result for this relatively small number of pulsars, sets of 25 values were drawn randomly from a sinusoidal distribution and the KS test was repeated on each. Of 100,000 sets of values, only 29 returned KS test results lower than 0.037\%, indicating a low probability that the low KS test value noted for the observed distribution occurred by chance. }.

The distribution would be affected by, for example, a systematic bias in the favoured positions of the fiducial planes. These were judged using the profile shape assuming the core-cone model \citep{ran93}. However, the alternative to this assumption, the `patchy' beam model suggested by \cite{lm88}, would also not explain the observed bias in $\alpha$. In this model the emission is generated randomly across the polar cap, meaning that there should be no systematic bias in the illumination of the beam with respect to the fiducial plane and hence no systematic bias in $\phi_{\mathrm{fid}}$. For the possible interpretations of the observed $\alpha$ distribution, see \cite{rwj14b}.

\subsection{Dependence on profile morphology}
\label{SectSinglesVsDoubles}

\cite{jw06} (JW06) noted that in their sample of young pulsars with high rates of rotational energy loss (comparable to our sample in terms of spin parameters, but comprising only 14 objects), double-peaked profiles were often brighter in the trailing components. Here we take this argument a step further, considering the case in which the asymmetry is sufficient that the leading component is unobserved. In $\S$~\ref{SectResults} we argued that this could apply to PSRs~J1730--3350, J1801--2451 and J1835--1106. This raises the question of whether all apparent single-component profiles in our sample are in fact such highly asymmetric doubles.

This pronounced version of the JW06 effect would strongly affect the estimated position of the fiducial plane. The profile appears as a single component and so the fiducial plane would in most cases be wrongly assumed to be at the peak. However, if in reality there is a missing leading component the true location of the fiducial plane will be close to the leading edge of the observed component. By not accounting for this effect where it is present, we will have overestimated $\phi_{\mathrm{fid}}$ and hence underestimated the offset in rotational phase between the fiducial plane and inflection point, affecting the determination of $\alpha$. Multiple-component profiles will not have been affected in this way, as even if an asymmetry is present both components are still discernible.

The fiducial plane position affects the measured $\alpha$ value 
in two ways. Firstly, the phase difference between the fiducial 
plane and inflection point is related to the derived emission height and as a consequence the beamwidth via 
Eqs.~\ref{EqBeamHalfOpeningAngle} and \ref{EqThetaPC}. A later fiducial 
plane implies a smaller beamwidth and it can be seen from 
Eq.~\ref{EqRhoContours} that if the beamwidth is underestimated 
$\alpha$ will also be underestimated for a given $W_{\mathrm{open}}$, 
the pulse longitude range covered by the open field lines. 
However, the fiducial plane 
position also affects the estimated $W_{\mathrm{open}}$, which is determined 
by the offset of the pulse edge furthest away from the fiducial plane. 
If $W_{\mathrm{open}}$ is underestimated the measured $\alpha$ 
value corresponding to a particular beamwidth will be overestimated. 

These two competing effects mean that adjusting the position of the 
fiducial plane has a complicated effect on the measured $\alpha$ value for a given pulsar. If all apparently single-component pulsars in our sample are subject to the JW06 effect, the systematic misplacement of the fiducial plane could potentially lead to a further skew towards low $\alpha$ values in the distribution of the single-component pulsars relative to that of the multiple-component pulsars. If we assume that the profile morphology is not $\alpha$-dependent the two distributions should be intrinsically similar, and hence the additional skew applicable to single-component pulsars would lead to a difference between our derived distributions. It is therefore desirable to investigate whether such an effect is indeed present in our sample, so that any affected $\alpha$ values could then be corrected.

\begin{figure}
\centering
\includegraphics[height=\hsize,angle=270]{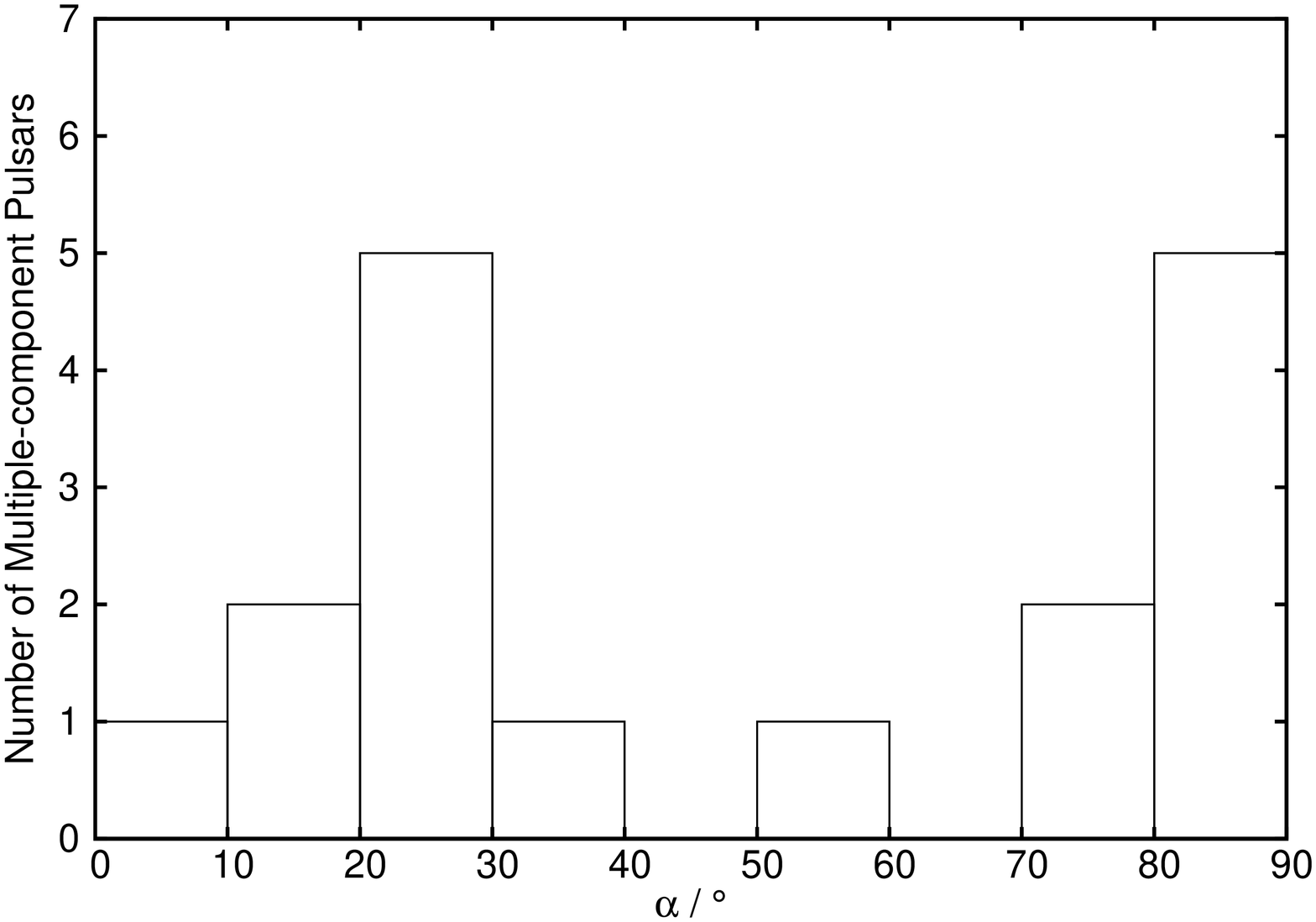}
\includegraphics[height=\hsize,angle=270]{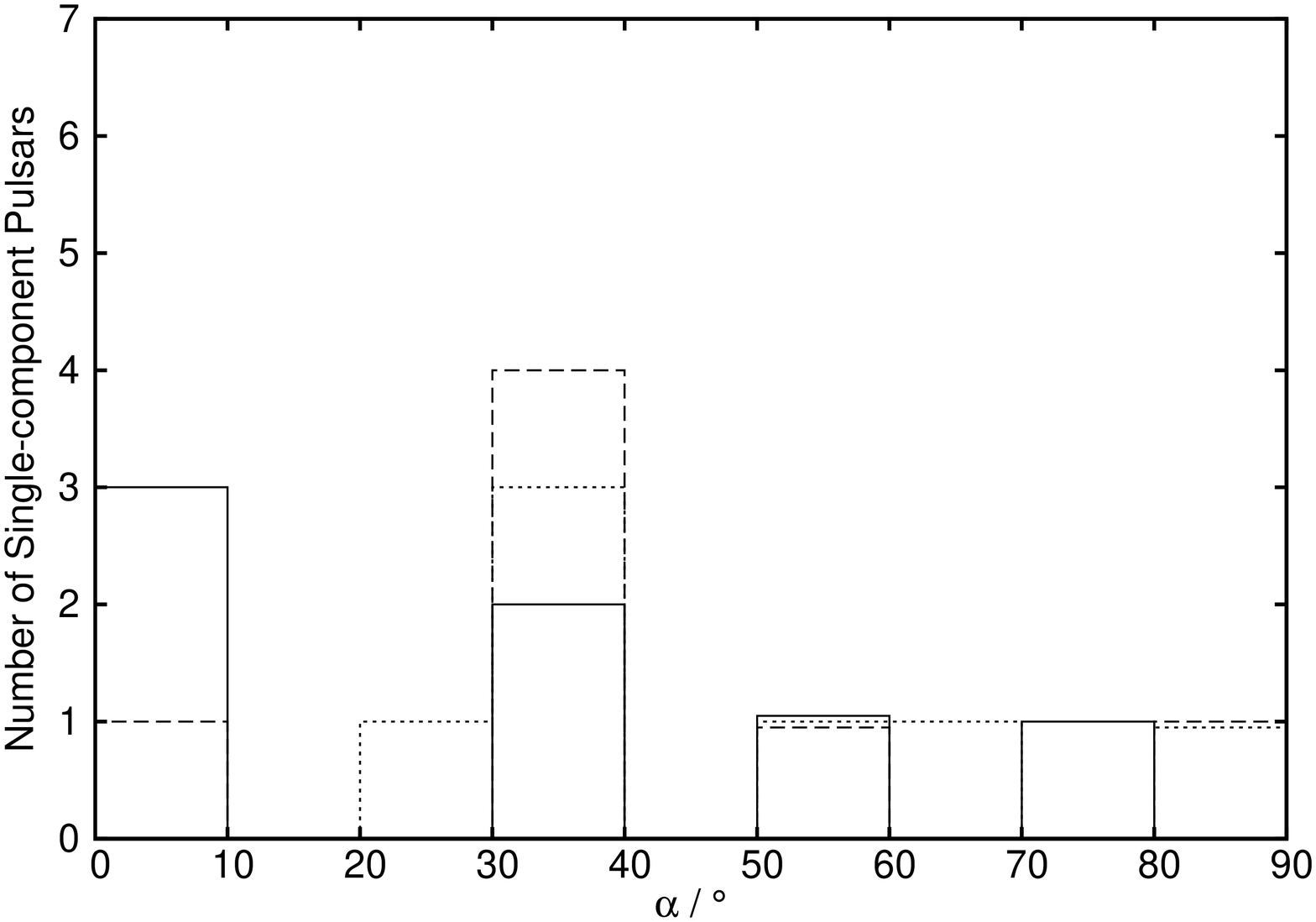}
\caption{\label{FigSinglesVersusMultiples} (Upper panel) Distribution of most likely $\alpha$ values of pulsars with profiles exhibiting multiple components. (Lower panel) Distribution of most likely $\alpha$ values of the single component pulsars when the fiducial plane is taken to be at the peak of the observed component (solid line), after applying the Johnston \& Weisberg effect to those pulsars (PSRs~J1730--3350, J1801--2451 and J1835--1106) for which there is a suggestion of a missing leading component (dashed line) and after making the assumption that each pulsar has an unobserved leading component (dotted line). For clarity the lines have been displayed with slight offsets in the bins where they would otherwise coincide. }
\end{figure}

To test this hypothesis we first of all calculated the $\alpha$ distribution of the single-component pulsars\footnote{Taken to be PSRs J0659$+$1414, J1357--6429, J1410--6132, J1709--4429, J1730--3350, J1801--2451 and J1835--1106.} when the fiducial plane of each was positioned at the profile peak. This distribution is shown as the solid line in the lower panel of Fig.~\ref{FigSinglesVersusMultiples} and can be compared with that derived for the multiple-component pulsars (upper panel of the same figure). The peak is later than the inflection point for PSRs~J1730--3350 and J1801--2451, resulting in a negative derived emission height and meaning that the favoured contour cannot be determined. However, the overall constraint on $\alpha$ (see Figs.~\ref{FigJ1730-3350} and \ref{FigJ1801-2451}) indicates that both these pulsars have small magnetic inclinations. For this reason both were assigned a value $\alpha = 0\degree$. PSR~J1513--5908 also yields a negative derived emission height in this situation. However, $\alpha$ is unconstrained for this pulsar (see Fig.~\ref{FigJ1513-5908}) and so no reliable estimate can be made in the absence of the favoured contour. For this reason, the pulsar was excluded from this part of the analysis. 

Applying the KS test to the alpha distributions derived for the multiple-component pulsars and the single-peaked profiles under the assumption that the JW06 effect is absent for all pulsars results in a probability of 24\%. Hence, there is no evidence for the distributions to be different.

The test was repeated using the $\alpha$ distribution (dashed line in the lower panel of Fig.~\ref{FigSinglesVersusMultiples}) resulting when the fiducial plane was placed at the leading edge of the pulse for the pulsars for which there was a suggestion of a missing leading component (PSRs~J1730--3350, J1801--2451 and J1835--1106). This therefore corresponds to the preferred values as they appear in Table~\ref{TabUncorrectedViewingGeometries}. The result of the KS test in this case was 57\%. This shows that applying the JW06 effect to these pulsars has a non-significant effect on the distribution. The test was also repeated with the fiducial plane positioned at the leading edge for all the single-component pulsars (dotted line in the lower panel of Fig.~\ref{FigSinglesVersusMultiples}), corresponding to the scenario that all single-peaked profiles have missing leading components. The KS test result was again 57\%. 

Applying the JW06 effect to some pulsars might increase the similarity between the distributions for single- and multiple-component pulsars, indicating that the effect may be present in some particular cases (and indeed for PSRs~J1730--3350 and J1801--2451 it appears necessary in order to avoid negative derived emission heights). However, the sample size is too small to draw any significant conclusions. Applying the effect to all single pulsars has little effect on the overall $\alpha$ distribution. We therefore take a conservative option, whereby we apply the effect in those cases for which there is evidence, but do not apply the effect to other single-component pulsars. The values given in Table~\ref{TabUncorrectedViewingGeometries} correspond to this situation.

There are reasons to expect that the JW06 effect is not applicable to all pulsars exhibiting a single-component profile. Firstly, PSR~J1119--6127 has a single-peaked profile in almost every observation. However, \cite{wje11} showed a transient component trails the often-seen component. The two RRAT-like components detailed in that paper are believed to frame the two main components, strongly suggesting the fiducial plane to be located after, not before, the `single' component. This then appears to be a case where the JW06 effect is not applicable.

Furthermore, the intensity asymmetry discussed in JW06 does not appear for all $\gamma$-ray-loud pulsars exhibiting double-component radio profiles. For example, the profile of PSR~J1105--6107 ($\S$~\ref{J1105}) shows two components with approximately equal peak intensities. Given that some double-component profiles do not show this asymmetry, it is plausible that there also exist some single profiles which are not subject to the effect and hence are intrinsically single, rather than extremely asymmetric double-component profiles.

\subsection{Comparison with constraints from $\gamma$-ray models}
\label{SectPierbattistaComparison}

Recently, \cite{phg+14} (henceforth PHG14) have derived maps 
of the `likelihood' of a 
given viewing geometry from $\gamma$-ray and radio profiles 
and hence determined the most likely emission geometry for a large
number of $\gamma$-ray pulsars based on four $\gamma$-ray models: the polar
cap \citep{mh03}, slot gap \citep{mh04a}, outer gap \citep{crz00} and one-pole caustic model \citep{rw10}. Since we also studied $\gamma$-ray
detected pulsars, PHG14 includes most of the pulsars examined in this
paper with the exception of PSR J1531--5610. The emission geometries
derived by PHG14 are based on very different information compared to
our study (e.g. radio polarization was not considered). It is
therefore interesting to investigate whether our measurements are
consistent with the geometries needed to explain $\gamma$-ray light
curves. Since $\beta$ is confined to a small range for radio-detected
pulsars and its precise value can be expected to be highly dependent
on details in the way the radio profiles are interpreted, we choose to
investigate if there is a correlation between the $\alpha$ values presented
in this paper and those by PHG14.

\begin{figure*}
\centering
\includegraphics[height=0.4\hsize,angle=270]{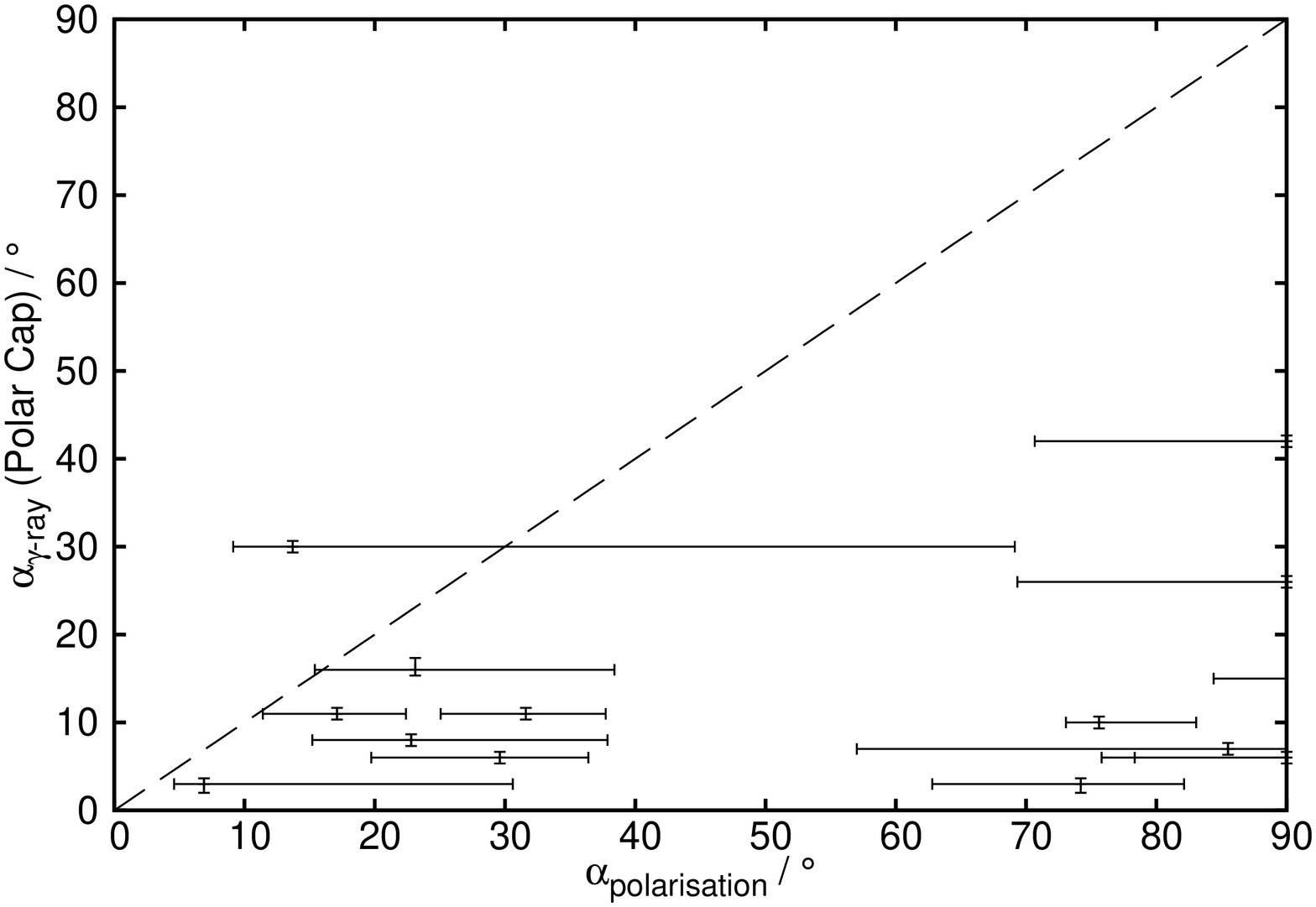}
\includegraphics[height=0.4\hsize,angle=270]{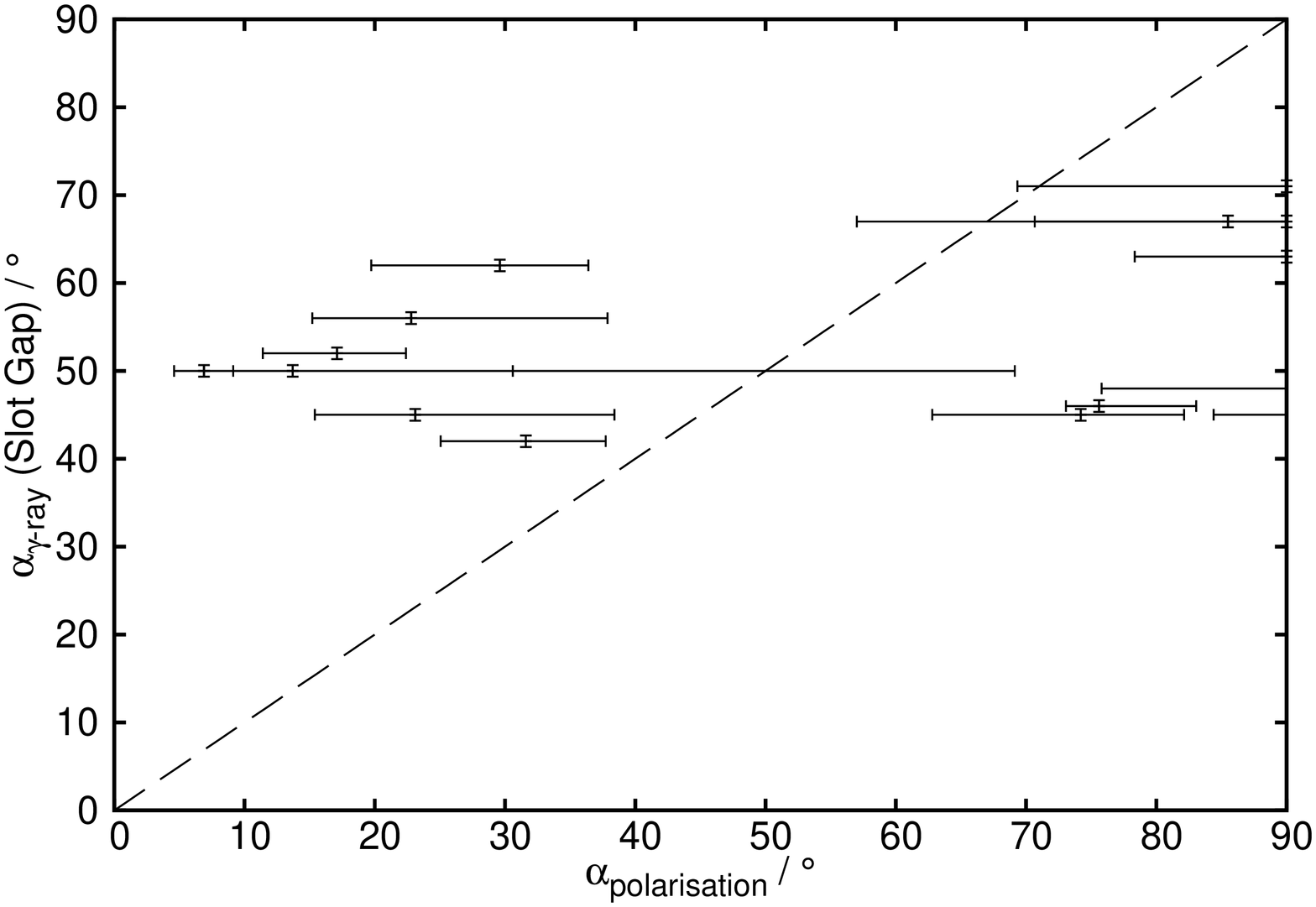}
\includegraphics[height=0.4\hsize,angle=270]{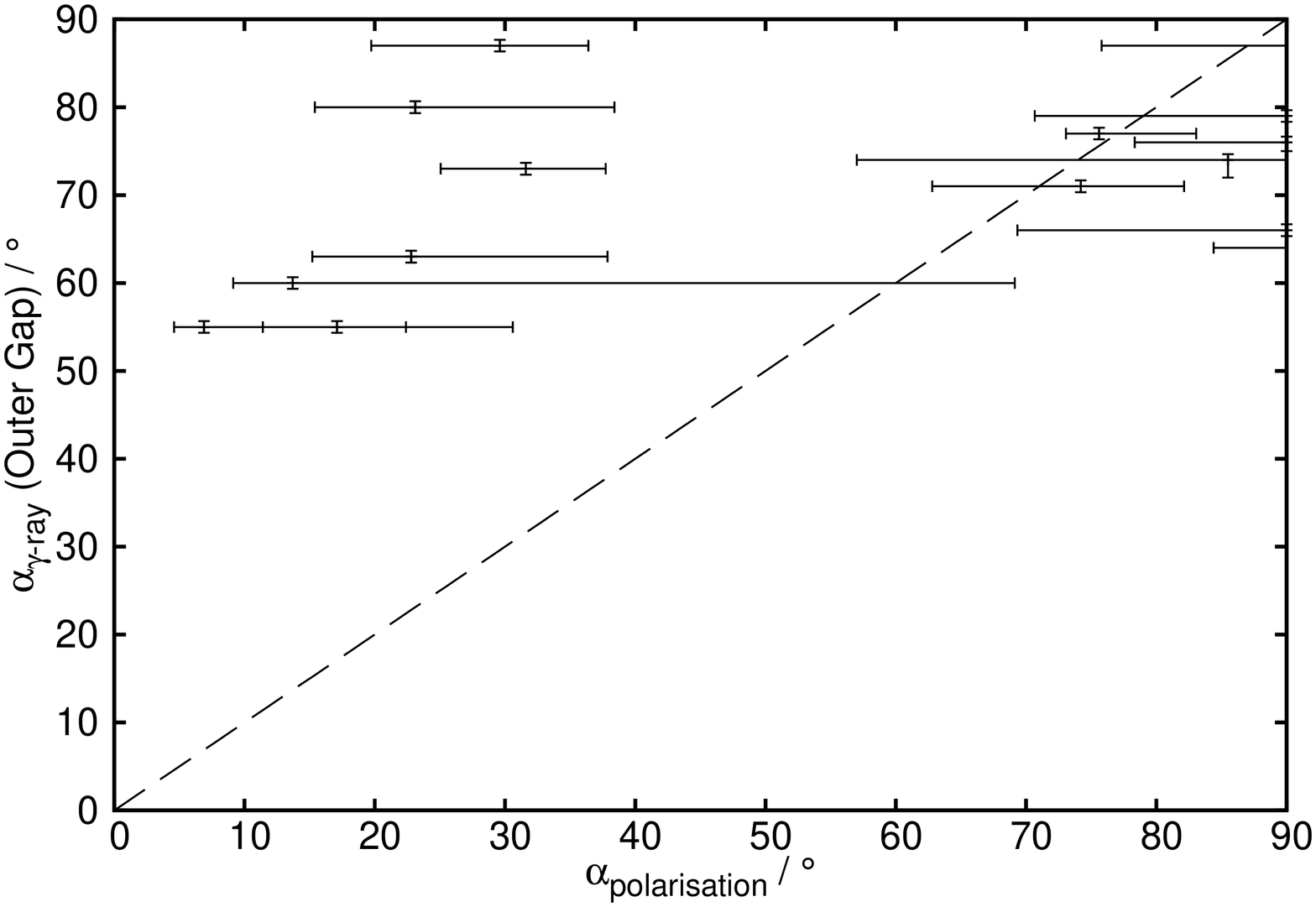}
\includegraphics[height=0.4\hsize,angle=270]{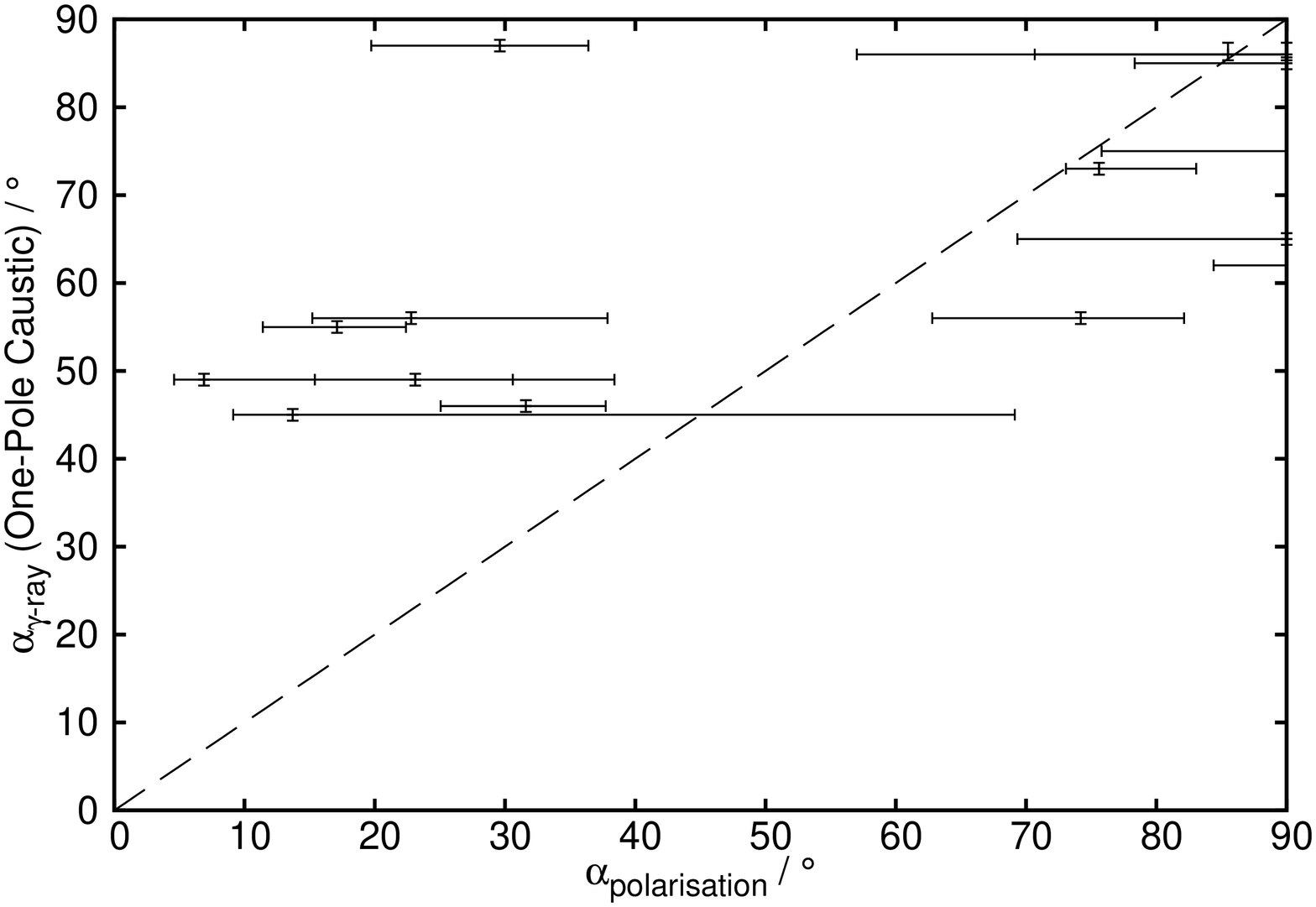}
\caption{\label{FigUsVsPierbattista} Comparison between $\alpha$ values derived in $\S$~\ref{SectResults} ($\alpha_{\mathrm{polarisation}}$) and by \protect \cite{phg+14} ($\alpha_{\gamma\mathrm{-ray}}$). The 1$\sigma$ errorbars are shown. The $\gamma$-ray emission models used by the latter were the polar cap (upper left), slot gap (upper right), outer gap (lower left) and one-pole caustic models (lower right). The diagonal line on each plot represents agreement between the two sets of values. }
\end{figure*}

There is inherent disagreement between the viewing geometries implied by the various $\gamma$-ray models. For instance, the relatively small emission region of the polar cap model, coupled with the typically wide $\gamma$-ray intensity profiles, implies low $\alpha$ values. In contrast to this, the other models propose the emission to be generated in the outer magnetosphere, resulting in larger derived magnetic inclination angles. This can be seen in Fig.~\ref{FigUsVsPierbattista}, which shows the $\alpha$ values derived by PHG14 against the values derived in $\S$~\ref{SectResults}. None of the four models can explain the large range of $\alpha$ values we have found in this paper. The low $\alpha$ values in particular are incompatible with the slot gap, outer gap and one-pole caustic models.

It can also be seen from Fig.~\ref{FigUsVsPierbattista} 
that the two sets of values are uncorrelated for all four 
$\gamma$-ray models considered by PHG14, in the sense that a large 
$\alpha$ value derived in this paper does not necessarily correspond 
to a large $\alpha$ value according to PHG14, although the 
values are subject to large errors. To investigate whether the two sets 
of values could be considered correlated within these errors a series 
of trials was performed. For each trial, two $\alpha$ values 
(corresponding to the constraints presented in this paper and 
to those presented by PHG14) were drawn randomly for each pulsar, 
presuming a Gaussian distribution function with widths 
determined by the quoted errors. 
The Spearmann rank-order correlation coefficient was then found 
between the two resulting sets of values. None of the four cases 
showed a significant tendency towards correlation.

There are several reasons why the two sets of values might be
uncorrelated. Firstly, it is possible that none of the $\gamma$-ray
models are a good description of the physics. If true, this might be
somewhat surprising given that the shape of the $\gamma$-ray
light-curves, at least for the models which place the extended
emission region in the outer magnetosphere, is in first order
determined by the pulse phase ranges where caustic emission is
expected to happen. One could expect this geometric effect to be at
some level independent of the detailed physics.

Another possibility is that the absence of a correlation between the
$\alpha$ values derived in this paper and by PHG14 is caused by the
fact that the interpretation of the radio data is very different. For
example, PHG14 only considered the shape of the radio profile and
assumed that the radio emission height follows an empirical relation
described by \cite{kg98}, whereas in this paper we considered both the
radio profile morphology and polarisation allowing us to
determine the emission height. Although these methods will be
inconsistent up to some level, it remains to be seen whether this
difference would have a significant effect on the derived $\alpha$
values from joint radio and $\gamma$-ray fitting. As shown by PHG14,
the effect of considering radio data is effectively to down-weight
solutions with large $\beta$ values. Since the different methods to
interpret the radio data will have a similar effect, it could be
expected that the fine details of the radio model only have a minor
effect on the resulting $\alpha$ values. How big this effect is can
potentially be determined by considering the constraints derived in
this paper together with the fits of $\gamma$-ray data.

\section{\label{SectConclusions}Conclusions}

In this paper we have presented constraints to the viewing geometry (characterised by the magnetic inclination angle, $\alpha$, and the impact parameter of the line of sight relative to the magnetic axis, $\beta$) for a sample of $\gamma$-ray-loud pulsars. This constraint has two components. The first is associated with the goodness of fit between the Rotating Vector Model and the observed PA curve. The second is associated with the emission beam half-opening-angle derived from the relative offset in pulse phase between the inflection point of the PA curve and the location of the fiducial plane in the profile (which represents the point of closest approach between the line of sight and the magnetic axis). In determining the latter constraint, we have taken into account uncertainties on the fiducial plane, inflection point and pulse edges in a systematic way. This new approach is more conservative and therefore in many cases allows regions of the ($\alpha$, $\beta$) parameter space to be objectively excluded with greater confidence. In addition to this conservative constraint, we also determined our preferred viewing geometry for each pulsar, allowing a statistical analysis of the results.

We find that the $\alpha$ distribution exhibits an unexpected skew towards low values. These pulsars are all younger than the various observationally estimated timescales for alignment of the magnetic inclination angle due to magnetic torques, suggesting the observed distribution should be the birth distribution for this sample. If the magnetic axis is randomly orientated with respect to the rotation axis at the birth of a neutron star, it follows that this birth distribution of $\alpha$ should be sinusoidal. Further to this, the pulsars investigated in this paper have all been detected in $\gamma$-rays. High energy models predict that pulsars are more easily detectable in $\gamma$-rays when the magnetic inclination is large, as the intensity modulation is more pronounced. This would be expected to introduce a selection effect resulting in an abundance of highly inclined pulsars relative to the sinusoidal distribution. These considerations make the observed tendency towards low $\alpha$ values surprising. This skew and its possible causes will be investigated in detail in \cite{rwj14b}. 

We have also reported evidence for absent leading components in some of the pulse profiles which exhibit only a single component. An argument can be made that these are extreme cases of the intensity asymmetry noted in double-component profiles of young pulsars by \cite{jw06}. However, the sample size is too small to determine whether this effect applies to all apparently single-component young pulsars.

We have found a lack of correlation between the $\alpha$ values derived here from radio intensity and polarisation data and those derived by \cite{phg+14} from consideration of the radio and $\gamma$-ray light curves. This indicates a possible problem in the interpretation of the radio and/or $\gamma$-ray data which requires further investigation. However, despite this uncertainty in $\alpha$, the tight constraints on $\beta$ which we have presented should prove useful for future attempts to determine the $\gamma$-ray emission mechanism.

\section*{Acknowledgments}

We wish to thank the referee Aris Karastergiou for his constructive input.    
Pulsar research at JBCA is supported by a Consolidated Grant from the UK
Science and Technology Facilities Council (STFC). The Parkes radio telescope 
is part of the Australia Telescope which is funded by the Commonwealth of 
Australia for operation as a National Facility managed by CSIRO.

~

\bibliographystyle{mn2e}

\bsp

\label{lastpage}

\end{document}